\documentclass[preprint,12pt]{elsarticle}
\usepackage[hidelinks]{hyperref}
\usepackage[hyperref,colornames,dvipsnames]{xcolor}

\usepackage{amssymb}
\usepackage{listings}
\usepackage{color}
\usepackage{amsmath}
\usepackage{booktabs} 
\usepackage{tabularx}

\usepackage{amsmath}
\usepackage{mathtools}
\usepackage{makeidx}
\usepackage{graphicx}
\usepackage{caption}
\usepackage{subcaption}
\usepackage[utf8]{inputenc}
\usepackage{feynmp-auto}
\usepackage{graphicx}

\biboptions{numbers,sort&compress}
\definecolor{lightgrey}{gray}{0.92}
\def\btab#1\etab{\begin{tabular}{p{75mm}p{75mm}}#1\end{tabular}}
\def\btabx#1\etabx{\begin{tabular}{p{60mm}p{50mm}}#1\end{tabular}}
\def\btaby#1\etaby{\begin{tabular}{p{15mm}p{95mm}}#1\end{tabular}}
\def\bcen{\begin{center}}
\def\ecen{\end{center}}
\def\bgfb#1\egfb{\bcen\centerline{\begin{tabularx}{1.2\linewidth}{|l
                X|}\hline#1\\\hline\end{tabularx}}\ecen}
\def\bgfbx#1\egfbx{\bcen\fcolorbox{black}{lightgrey}{\parbox{118mm}{\btabx#1\etabx}}\ecen}
\def\bgfbalign#1\egfbalign{\bcen\fcolorbox{black}{lightgrey}{\parbox{118mm}{\btaby#1\etaby}}\ecen}
\def\beq{\begin{equation}}
\def\eeq{\end{equation}}
\def\bea{\begin{eqnarray}}
\def\eea{\end{eqnarray}}
\newcommand{\ihixs}{\texttt{iHixs}}
\newcommand{\spart}{\widehat{\sigma}}

\def\blst#1\elst{\begin{lstlisting}#1\end{lstlisting}\vspace{-1.5\baselineskip}}

\definecolor{mygreen}{rgb}{0,0.6,0}
\definecolor{mymauve}{rgb}{0.58,0,0.82}
\hypersetup{
    colorlinks
}
\journal{CPC}
\newcommand{\fakecaption}{%
    \refstepcounter{table}%
}

\begin{document}
\hypersetup{
    colorlinks,
    linkcolor=[rgb]{0.15,0.35,0.65},
    citecolor=[rgb]{0.15,0.35,0.65},
    urlcolor=[rgb]{0.15,0.35,0.65}
}
\begin{frontmatter}


\tnotetext[]{CERN-TH-2018-019, SLAC-PUB-17222}

\title{iHixs 2 - Inclusive Higgs Cross Sections}

\author[label1]{Falko Dulat}
\ead{dulatf@slac.stanford.edu}
\author[label2]{Achilleas Lazopoulos}
\ead{lazopoli@phys.ethz.ch}
\author[label3]{Bernhard Mistlberger}
\ead{bernhard.mistlberger@gmail.com}

\address[label1]{SLAC National Accelerator Laboratory, Stanford University, Stanford, CA 94039, USA}
\address[label2]{Institute for Theoretical Physics, ETH Z\"urich, 8093 Z\"urich, Switzerland}
\address[label3]{CERN Theory Division, CH-1211, Geneva 23, Switzerland}

\begin{abstract}
We present a new release of the program \ihixs. 
This easy-to-use tool allows to derive state of the art predictions for the inclusive production cross section of a Higgs boson at hadron colliders in the gluon fusion production mode. 
This includes the most up-to-date corrections in perturbative QCD and electro-weak theory, effects due to finite quark masses as well as an option to perform threshold resummation. 
In particular, exact perturbative QCD corrections through N$^3$LO are included in the heavy top quark effective theory.
Furthermore, \ihixs{} contains automatic routines that allow to assess residual uncertainties on the prediction for the Higgs boson production cross section according to well established standard definitions.
\ihixs{} can be obtained from \url{https://github.com/dulatf/ihixs}.
\end{abstract}





\end{frontmatter}


\newpage
\noindent {\bf PROGRAM SUMMARY}                                              \\
  \begin{small}
  {\bf Manuscript Title:} iHixs 2 --   Inclusive Higgs Cross Sections        \\
  {\bf Authors:}
  Falko Dulat, Achilleas Lazopoulos, Bernhard Mistlberger                    \\
  {\bf Program Title:} iHixs                                                 \\
  {\bf Journal Reference:}                                                   \\
  {\bf Catalogue identifier:}                                                \\
  {\bf Licensing provisions:} None.                                          \\
  {\bf Programming language:} C++.                                           \\
  {\bf Computer:} Platforms on which LHAPDF 6 and a C++11 compatible compiler are available.                                                                   \\
  {\bf Operating system:} Linux, MacOS.                                      
  {\bf Classification:} 11.1 General, High Energy Physics and Computing.     \\ 
  {\bf External routines/libraries:} Cuba, Chaplin (shipped with code)       \\
  {\bf Nature of problem:} Determine the inclusive Higgs boson cross section at hadron colliders in the gluon fusion production mode.                                     \\
  {\bf Solution method:}  Numerical convolution of analytic partonic cross sections with parton distribution functions.                                     \\
  {\bf Restrictions:}                                                        \\
  {\bf Running time:} Several seconds to minutes.                                          \\
\end{small}

\newpage

\section{Introduction}
The discovery of the Higgs boson at the Large Hadron Collider (LHC) at CERN by ATLAS~\cite{Aad2012} and CMS~\cite{Chatrchyan2012} heralded the beginning of the age of Higgs boson measurements.
The newly found boson represents a window into an entirely new sector of particle physics. 
The exploration of the properties of the Higgs boson sheds light on its nature and provides a potent tool for the investigation of possible physics beyond the Standard Model (SM) of particle physics.
The rapidly increasing amount of collected data lead to a swift transition from discovery to precision measurements of the features of the Higgs boson.
Attributes like the mass, the spin or the parity of the newly found boson have been determined already to astounding levels of precision and seem in remarkable agreement with the SM.

One of the most essential observables in Higgs boson phenomenology is the probability to produce a Higgs boson in the collision of protons. 
This quantity allows on its own for a stringent test of the SM and is key for the extraction of coupling constants. 
In this article we present a numerical tool, \ihixs{}, that allows to predict the inclusive production cross section of a Higgs boson at a hadron collider.
Specifically, we focus on the dominant mechanism to produce a Higgs boson: gluon fusion. 
The explicit aim of this article is to unite all state of the art contributions to the inclusive Higgs boson production cross section in a single numerical code.
The theoretical foundation of this program was presented in ref.~\cite{Anastasiou:2016cez} that includes a critical assessment of all contributions and their respective uncertainties (see also refs.~\cite{deFlorian:2016spz,Harlander:2016hcx}).

The Born level cross section for the production of a Higgs boson through the fusion of two gluons via a top quark loop was derived long ago in ref.~\cite{Wilczek1977}.
Perturbative corrections to the leading order (LO) cross section were subsequently discovered to be sizeable. 
The largest effect is due to QCD corrections. 
Such corrections can in a first approximation be computed in an effective theory (EFT) where the top quark is considered to be infinitely heavy~\cite{Chetyrkin:1997un,Schroder:2005hy,Chetyrkin:2005ia,Kramer:1996iq}. 
EFT corrections were computed at next-to-LO (NLO) in ref.~\cite{Dawson:1990zj} , at next-to-next-to-LO (NNLO) in refs.~\cite{Anastasiou2002,Harlander:2002wh,Ravindran:2003um} and at next-to-next-to-next-to-LO (N$^3$LO) in refs.~\cite{Anastasiou:2015ema,Anastasiou:2016cez,Mistlberger:2018}. 
In order to achieve predictions at the level of precision required for the comparison with experimental measurements it is of paramount importance to improve pure effective theory predictions through the inclusion of effects due to finite quark masses. 
NLO QCD corrections in the full SM were computed in refs~\cite{Graudenz:1992pv,Spira:1995rr}. Beyond NLO only approximate results in terms of a power series of the cross section in inverse powers of the top quark mass are available at NNLO~\cite{Harlander:2009my,Pak:2009dg}. 
Corrections due to electro-weak effects were computed in refs.~\cite{Actis:2008ts,Actis:2008ug,Aglietti:2004nj,Bonetti:2016brm} and even mixed QCD-electro-weak effects were approximated in refs.~\cite{Anastasiou:2008tj,Bonetti:2018aa}.
\ihixs{} combines all the above effects in one single tool and allows to study their impact on the inclusive production probability of a Higgs boson at the LHC in detail.

Until recently, N$^3$LO QCD corrections derived in the heavy top quark EFT were based on a so-called threshold expansion of the partonic cross section.
Recently, exact results for these N$^3$LO cross sections became available~\cite{Mistlberger:2018} and we include them for the first time in a numerical code that allows to derive predictions for LHC phenomenology.
In particular, this allows us to further improve on the prediction of ref.~\cite{Anastasiou:2016cez} for the Higgs boson cross section and we update the current state of the art prediction.

In order to derive reliable predictions for LHC phenomenology a critical appraisal of residual uncertainties on the Higgs boson production cross section is vital. 
A careful analysis of such sources of uncertainty was carried out in ref.~\cite{Anastasiou:2016cez}
and \ihixs{} includes automatic routines that follow the prescriptions outlined therein to quantify
these uncertainties.
We identify as sources of uncertainty the truncation of the perturbative QCD and electro-weak expansion, the approximation of finite quark mass effects and the imprecise knowledge of the value of coupling constants and parton distribution functions.  

While previous versions of \ihixs~\cite{Anastasiou:2011pi,Anastasiou:2012hx} allowed already to derive predictions for the inclusive Higgs boson production cross section the new version presented in this article is distinct in several new features.
The heavy top quark EFT QCD corrections are now included exactly through N$^3$LO,
threshold resummation can be performed automatically through N$^3$LL using classical QCD techniques~\cite{Bonvini:2014joa,Catani:2014uta,Anastasiou:2016cez} or soft-collinear effective theory (SCET)~\cite{Ahrens:2008nc,Ahrens:2008qu}, and uncertainties of the Higgs boson cross sections can be automatically assessed according to the standards defined in refs.~\cite{deFlorian:2016spz,Anastasiou:2016cez}.

This article is structured as follows.
In section~\ref{sec:setup} we introduce the main definitions of the ingredients of \ihixs{} and explain them in some detail. 
Next, in section~\ref{sec:Uncertainties} we discuss sources of residual uncertainties on the Higgs boson cross section and outline how they are estimated in \ihixs.
In section~\ref{sec:pheno} we derive state-of-the-art phenomenological predictions for the Higgs boson production cross section at hadron colliders.
Subsequently, we present a detailed manual of \ihixs{} in section~\ref{sec:manual}. 
Finally, we conclude in section~\ref{sec:conclusions}.

\section{Set-Up}
\label{sec:setup}
In this article we present the numerical tool \ihixs{} that allows for the computation of the probability to produce a Higgs boson in the collision of protons via the gluon fusion production mode
\begin{equation}
    {\rm Proton}(P_1) + {\rm Proton}(P_2) \to H(p_h) + X\,.
\end{equation}
$P_1$ and $P_2$ are the momenta of the colliding protons and $p_h$ the momentum of the Higgs boson.
In collinear factorization, the hadronic Higgs boson production cross section can be written as
\beq
\label{eq:xsdiffhad2}
\sigma_{PP\rightarrow H+X}(\mu_R,\mu_F)=
\tau \sum_{i,j} \int_\tau^1 \frac{dz}{z}\int_{\frac{\tau}{z}}^1 \frac{dx_1}{x_1} f_i(x_1,\mu_F)f_j\left(\frac{\tau}{x_1 z},\mu_F\right)
\frac{1}{z}\spart_{ij}(\mu_F,\mu_R).
\eeq
Here, we factorize long and short range interactions into parton distribution functions $f_i(x)$ and
partonic cross sections $\spart_{ij}$. 
The momenta of the colliding partons are related to the proton momenta through the momentum
fractions $x_i$ as $p_1=x_1 P_1$ and $p_2=x_2 P_2=\frac{\tau}{x_1 z}P_2$. 
We define
\bea
\tau&=&\frac{m_h^2}{S}\,,\hspace{1cm} S=(P_1+P_2)^2\,,\nonumber\\
z&=&\frac{m_h^2}{s}\,,\hspace{1cm} s=(p_1+p_2)^2\,.
\eea
The sum over $i$ and $j$ ranges over all contributing partons. Furthermore, we define the variable $\bar z= 1-z$.
The partonic cross section $\spart$ depends on the factorization scale $\mu_F$ and the
renormalization scale $\mu_R$. 

The parton distributions are extracted from experimental measurements by various
groups~\cite{Dulat:2015mca,Harland-Lang:2014zoa,Ball:2014uwa,Alekhin:2016uxn,Butterworth:2015oua} and
are accessed in our program via the \texttt{LHAPDF} framework~\cite{Buckley:2014ana}.
Our partonic cross sections include a large variety of effects that combined allow for the currently most precise prediction of the inclusive Higgs boson production cross section.

Let us begin by defining our master formula for the partonic cross section before we explain it in detail later.
\bea
\label{eq:masterdec}
\hat \sigma_{ij}&=& \textcolor{BlueViolet}{\text{R}_{\text{LO}} C^2} \left[ \sigma_{ij}^{\text{LO},\,\,\text{EFT}}+\sigma_{ij}^{\text{NLO},\,\,\text{EFT}}+ \sigma_{ij}^{\text{NNLO},\,\,\text{EFT}}+\sigma_{ij}^{\text{N$^3$LO},\,\,\text{EFT}}\right]\nonumber\\
&+& \delta \sigma_{ij}^{\text{LO},\,\,\text{(t,b,c)}}+\delta
\sigma_{ij}^{\text{NLO},\,\,\text{(t,b,c)}}+\delta \sigma_{ij}^{\text{NNLO},\,\,\text{(t)}}+
\textcolor{BlueViolet}{\textrm{R}_{\textrm{LO}}C^2}\,\delta \sigma_{ij}^{\text{Res}}\,.\nonumber\\
\eea
We define the combined Wilson coefficient,
\bea
\label{eq:WCDef}
C&=&C_{\text{QCD}}+\lambda_{\text{EWK}}(1+\frac{\alpha_S}{\pi}C_{1w}+\dots).\nonumber\\
C_{\text{QCD}}&=&\sum\limits_{i=0}^3 \left(\frac{\alpha_S}{\pi}\right)^i C_{\text{QCD}}^{(i)}\,.
\eea
Here $C_{\text{QCD}}$ is the QCD Wilson coefficient, matching the heavy top quark EFT to QCD with
finite masses and $\lambda_{\text{EWK}}$ is an effective Wilson coefficient incorporating
electroweak corrections.
\ihixs{} enables the user to choose which of the contributions in eq.~\eqref{eq:masterdec} and eq.~\eqref{eq:WCDef} should be taken into account in cross section predictions.
In the following we will discuss the individual contributions. 

\subsection{Effective Theory}
Perturbative corrections in QCD are known to be large and thus of significant importance for hadron collider phenomenology. 
The gluon fusion production cross section is loop induced process and the computation of high order corrections is consequently rather difficult. 
A very successful strategy to approximate higher order QCD corrections is the computation of
perturbative corrections within an effective theory (EFT) where the top quark is considered to be infinitely heavy and all 
other quarks to be massless~\cite{Inami1983,Shifman1978,Spiridonov:1988md,Wilczek1977}. 
This effective theory is described by the Lagrangian density 
\beq
\mathcal{L}_{\text{eff}}=\mathcal{L}_{SM,5}+\frac{\alpha_S}{12 \pi v} C_{\text{QCD}} H G_{\mu\nu}^a
G_a^{\mu\nu}\,,
\eeq
where H is the Higgs field, $ G_{\mu\nu}^a $ is the gluon field strength tensor and $\mathcal{L}_{SM,5}$ denotes the SM Lagrangian with $n_f$ = 5 massless quark flavours. 
The Wilson coefficient $C_{QCD}$ is obtained by matching the effective theory to the full SM in the limit where the top quark is infinitely heavy~\cite{Chetyrkin:1997un,Schroder:2005hy,Chetyrkin:2005ia,Kramer:1996iq}. It is implmeneted in~\ihixs{} through three loops, in both the on-shell scheme as well as the $\overline{\textrm{MS}}$-scheme.
The corrections to the partonic cross section in the effective theory at NLO~\cite{Dawson:1990zj}
, at NNLO~\cite{Anastasiou2002,Harlander:2002wh,Ravindran:2003um} and at
N$^3$LO~\cite{Anastasiou:2015ema,Anastasiou:2016cez,Mistlberger:2018} are currently available and implemented in
\ihixs{}.

The partonic cross sections $\sigma_{ij}^{\text{N$^n$LO},\,\,\text{EFT}}$ in
eqn.~\eqref{eq:masterdec} correspond to the corrections obtained in this effective theory at order
$n$ after factoring out the Wilson coefficient $C_{\textrm{QCD}}$. 
Higher order corrections to the cross section, due to perturbative corrections to the Wilson
coefficient, can then be taken into account consistently by including the corrected Wilson
coefficient from eq.~\eqref{eq:WCDef}.
The leading order cross section in the effective theory is given by
\beq
\label{eq:effectiveLO}
\sigma_{ij}^{\text{LO},\,\,\text{EFT}}=\frac{\alpha_S^2}{72\pi v^2(n_c^2-1)} \delta(1-z)\,.
\eeq
Where, $n_c$ is the number of colours. The Dirac delta function $\delta(1-z)$ acts as a distribution on the parton distribution functions.

\subsection{Mass Effects at LO and NLO}
In the full standard model with finite quark masses, the leading order cross section in the gluon
fusion production mode, mediated by massive quark loops, is given by,
\beq
\label{eq:sigmaLOexact}
\sigma_{ij}^{\text{LO}}=\frac{\alpha_S^2}{72\pi v^2(n_c^2-1)} \left|\sum\limits_q Y_q \tau_q \frac{3}{2} A(\tau_q)\right|^2\delta(1-z)\,,
\eeq
with 
\beq
\tau_q=\frac{4m_q(m_q-i\Gamma_q)}{s}\,.
\eeq
Here, $m_q$ and $\Gamma_q$ are the mass and the width of the quark with
flavour $q$. 
The $Y_q$ are anomalous rescalings of the Yukawa couplings that are identically one in the Standard
Model, but might deviate from unity in beyond the Standard Model scenarios.
$A(\tau_q)$ is the famous loop factor defined as
\bea
A(\tau_q)=1-\frac{1}{4}\frac{(1+x_q)^2}{(1-x_q^2)}\log\left(x_q\right)^2\,,\hspace{1cm}x_q=\frac{-\tau_q}{(\sqrt{1-\tau_q}+1)^2}\,.
\eea
In the limit of infinite or vanishing quark mass we find that 
\beq
\lim\limits_{m_q\rightarrow \infty} \frac{3}{2}\tau_q A(\tau_q)=1\,,\hspace{1cm}\lim\limits_{m_q\rightarrow 0} \frac{3}{2}\tau_q A(\tau_q)=0\,.
\eeq
Our normalisation was chosen such that if we are sending the top quark mass to infinity and set all other quark masses to zero we reproduce the effective theory cross section at LO, eq.~\eqref{eq:effectiveLO}
\beq
\lim\limits_{m_t\rightarrow \infty ,\,\, m_{q\neq t}\rightarrow 0} \sigma_{ij}^{\text{LO}}=|Y_t|^2 \sigma_{ij}^{\text{LO},\,\,\text{EFT}}\,.
\eeq
In order to account for top-quark effects at LO in eq.~\eqref{eq:masterdec} we define the ratio
\beq
\text{R}_{\text{LO}}=\frac{\sigma_{ij}^{\text{LO},\,t}}{
    \sigma_{ij}^{\text{LO},\,\,\text{EFT}}}\,,
\eeq
The superscript $t$ indicates that the sum over quark flavours in eq.~\eqref{eq:sigmaLOexact} includes here only the top quark.
We rescale all higher order effective theory cross sections in eq.~\eqref{eq:masterdec} with this
ratio, defining the so-called \emph{rescaled effective theory} (rEFT).

Beyond the factorized corrections due to the finite top mass, at leading order, we also take into
account the exact dependence on the top, bottom and charm mass through the quantity
\beq
\delta \sigma_{ij}^{\text{LO},\,\,\text{(t,b,c)}}=\sigma_{ij}^{\text{LO},\,\,\text{(t,b,c)}}-\Bigl[C_{\text{QCD}}^2 \text{R}_{\text{LO}}\sigma_{ij}^{\text{EFT}}\Bigr]_{\alpha_S^2}\,,
\eeq
where the second term in the above equation, containing only terms proportional to $\alpha_S^2$, subtracts the LO rEFT contribution, in order to avoid double counting.
The effective theory cross section is defined as 
\beq
\sigma_{ij}^{\text{EFT}}=\sigma_{ij}^{\text{LO},\,\,\text{EFT}}+\sigma_{ij}^{\text{NLO},\,\,\text{EFT}}+ \sigma_{ij}^{\text{NNLO},\,\,\text{EFT}}+\sigma_{ij}^{\text{N$^3$LO},\,\,\text{EFT}}\,.
\eeq
The label $(t,b,c)$ indicates that we include corrections to the Higgs boson production cross section due to finite top, bottom and charm quark masses.

Exact QCD corrections to the gluon fusion cross section with full dependence on the quark masses are known at NLO~\cite{Graudenz:1992pv,Spira:1995rr}.
We take them into account, in analogy to the corrections at LO, by defining the correction to the effective theory cross section due to
quark masses at NLO,
\beq
\delta \sigma_{ij}^{\text{NLO},\,\,\text{(t,b,c)}}=\sigma_{ij}^{\text{NLO},\,\,\text{(t,b,c)}}-\Bigl[C_{\text{QCD}}^2 \text{R}_{\text{LO}}\sigma_{ij}^{\text{EFT}}\Bigr]_{\alpha_S^3}\,.
\eeq
Again, we need to subtract the rEFT corrections at the appropriate order to avoid double counting.
The mass dependent NLO correction is normalized such that it vanishes in the effective theory limit.

Due to the truncation of the perturbative series in the strong coupling constant at finite order,
our predictions depend on the choice of renormalization scheme. \ihixs{} offers the choice of the
two most commonly used schemes, the on-shell scheme, as well as the $\overline{\textrm{MS}}$
scheme and incorporates the Wilson coefficients, as well as the anomalous dimensions used for quark
mass evolutions in the both schemes.

\subsection{Mass Effects at NNLO}
Currently, corrections beyond NLO in exact QCD are unknown. 
In refs.~\cite{Harlander:2009my,Pak:2009dg} NNLO corrections were approximated by performing an expansion of the partonic cross section in $\frac{m_h}{m_t}$.
The NNLO corrections to the cross section can then be written as
\beq
\sigma_{ij}^{\text{NNLO}}=\sigma_{ij}^{\text{NNLO},\,\text{approx.}}+\mathcal{O}\left(\left(\frac{m_h^2}{m_t^2}\right)^4\right)\,.
\eeq
The numerically largest perturbative corrections arise due to contributions involving a gluon in the partonic initial state.
We include the approximate NNLO correction due to the top quark mass in the gluon-gluon and quark-gluon channel in our partonic cross section, eq.~\eqref{eq:masterdec}, as 
\beq
\delta
\sigma_{ij}^{\text{NNLO},\,\,\text{(t)}}=\sigma_{ij}^{\text{NNLO},\,\text{approx.}}-\Bigl[C_{\text{QCD}}^2
\text{R}_{\text{LO}}\sigma_{ij}^{\text{EFT}}\Bigr]_{\alpha_S^4}\quad\textrm{for}\quad(ij)\in\{(gg),(gq)\},
\eeq

\subsection{Electro-Weak Effects}
Corrections to the Higgs boson production cross section due to electro-weak physics are an important
ingredient for precision predictions.
The purely virtual leading corrections were computed in refs.~\cite{Actis:2008ts,Actis:2008ug,Aglietti:2004nj,Bonetti:2016brm}. 
In accordance with the \emph{complete factorisation} approach we include them in terms of a modification of our QCD Wilson coefficient.
To this end, we define the quantity $\lambda_{\text{EWK}}$ to be the ratio of the leading electro-weak corrections of ref.~\cite{Actis:2008ug} to the Born cross section and include it in eq.~\eqref{eq:WCDef}.

Corrections beyond LO in QCD and electro-weak physics are currently unknown. 
They were approximated in an effective theory of infinitely heavy W and Z bosons and top quark in ref.~\cite{Anastasiou:2008tj}.
In this approach the electro-weak gauge bosons are integrated out and calculations are performed in a framework where the QCD Wilson coefficient receives a modification.
The corrections in this approximation is taken into account in \ihixs{} by including the coefficient
\beq
C_{1w} =\frac{7}{6}
\eeq
in eq.~\eqref{eq:WCDef}. 
Recently, the mixed QCD-electroweak corrections were also approximated using the first term of a threshold expansion in ref.~\cite{Bonetti:2018aa}. 
The obtained results are in good agreement with the approach outlined above.

\subsection{Threshold Resummation}
Threshold resummation involves the summation of logarithms $\log(1-z)$ to all orders in the strong coupling expansions in the limit of $z\rightarrow 1$. 
In refs.~\cite{Bonvini:2014joa,Catani:2014uta,Anastasiou:2016cez} resummation of the Higgs boson cross section to N$^3$LL was presented in the conventional QCD resummation framework.
In refs.~\cite{Anastasiou:2016cez} resummation to N$^3$LL was performed in the framework of soft-collinear effective theory.
Both resummation frameworks have in common that they modify the cross section beyond the fixed order
accuracy, which we capture through the coefficient,
\beq
\delta \sigma_{ij}^{\text{Res}}=\text{R}_{\text{LO}}\Biggl(\sigma_{ij}^{\text{Res},\,\text{scheme}}-\Bigl[\sigma_{ij}^{\text{Res},\,\text{scheme}} \Big]_{\text{N$^n$LO}} \Biggr)=\mathcal{O}(\alpha_S^{n+1})\,.
 \label{eq:resum}\eeq
Here, $\sigma_{ij}^{\text{Res},\,\text{scheme}}$ is the threshold resummed cross section in a certain resummation scheme. 
The second term in the square bracket in the above equation corresponds to the threshold resummed cross section expanded to fixed order.
If we include $n$ orders in the fixed order expansion this guarantees that we modify our perturbative cross section only at the level of N$^{n+1}$LO corrections.
The first coefficient in eq~\eqref{eq:resum} is implemented in \ihixs{} in both traditional
soft-gluon resummation as well as through SCET resummation and the user can choose
to compute either one. 

Traditional soft gluon resummation operates in Mellin-space in order to factorize the various
contributions to the cross section. The cross section in eq.~\eqref{eq:xsdiffhad2} factorizes in
Mellin-space as,
\beq
\sigma(N) = \sum_{ij}f_{i}(N)f_j(N)\widehat{\sigma}_{ij}(N)\,,
\eeq
with the Mellin moments,
\beq
f_i(N) = \int_0^1\textrm{d}zz^{N-1}f_i(z)\,,\quad \widehat{\sigma}_{ij}
= \int_0^1\textrm{d}zz^{N-1}\frac{\widehat{\sigma}_{ij}}{z}\,.
\eeq
We invert the Mellin transform in \ihixs{} numerically by evaluating the integral,
\beq
\sigma^{\mathrm{Res,soft-gluon}}(\tau) = \sum_{ij}\int_{c-i\infty}^{c+i\infty}\frac{\textrm{d}N}{2\pi
    i}\tau^{1-N}f_i(N)f_j(N)\widehat{\sigma}_{ij}(N)\,.
\eeq
The resummation computes $\widehat{\sigma}_{ij}(N)$ by exponentiating the constant and
logarithmically divergent contributions in the limit $N\to\infty$ into the all-order resummation
formula~\cite{Sterman:1986aj,Catani:2003zt,Catani:1989ne,Catani:1990rp},
\beq
\widehat{\sigma}(N)_{ij}=\delta_{ig}\delta_{jg} \alpha_s^2 \sigma_0
C_{gg}(\alpha_s)\exp[\mathcal{G}_{H}(\alpha_s,\log(N))]\,.
\eeq
The matching coefficient $C_{gg}$ contains terms that are constant in the limit while
$\mathcal{G}_{H}$ exponentiates the large logarithms. It has been computed through N$^3$LL accuracy in terms
of the QCD $\beta$ function and the cusp anomalous dimension in
refs.~\cite{Moch:2005ba,Catani:2014uta,Bonvini:2014joa}.
In addition to the large logarithms, various resummation schemes, exponentiate terms that are
subleading in the limit $N\to\infty$, e.g.~by replacing $\log(N)\to\psi(N)$. While these schemes are
formally equivalent, they agree in the defining limit $N\to\infty$, they introduce numerical
differences due to subleading terms. Various schemes have been compared in
ref.~\cite{Anastasiou:2016cez} and \ihixs{} provides a selection for common choices of threshold
resummation schemes.

As opposed to traditional threshold resummation, resummation in the SCET framework is directly
performed in $z$ space. Here the cross section is factorized into a hard function $H$ and a soft
function $\tilde{S}$; the large threshold logarithms are then resummed by solving the
renormalization group equations for these operators.
Following refs.~\cite{Ahrens:2008nc,Ahrens:2008qu}, the cross section  was resummed in the SCET framework in
ref.~\cite{Anastasiou:2016cez}. Schematically it can be written as,
\beq
\sigma_{ij}^{\textrm{Res,SCET}} \propto \delta_{ig}\delta_{jg} \sigma_0 C^2(m_t^2,\mu_t^2) H^2(m_H^2,\mu^2,\mu_t^2,\mu_h^2,\mu_s^2) \tilde{S}(\mu_s) U(\mu^2,\mu_h^2,\mu_s^2,\mu_t^2)\,.
\eeq
Here $C$ is the Wilson coefficient defined in eq.~\eqref{eq:WCDef}. 
The function $U$ exponentiates the infrared structure and resums renormalisation group logarithms.
By continuing the logarithms of the hard scale $\mu_h$ to the space-like region, \ihixs{} optionally enables the resummation of $\pi^2$ terms.
We refer the reader to ref.~\cite{Anastasiou:2016cez} for details on the method.

In ref.~\cite{Anastasiou:2016cez} the impact of threshold resummation effects were studied and found to be small beyond N$^3$LO in perturbative QCD. 
As a consequence we consider them to be an important tool to study the impact of potential higher order corrections but do not include them in our default recommendation for cross section predictions.
However, \ihixs{} provides the option to include these effects.

\section{Uncertainties}
\label{sec:Uncertainties}
In the previous section we summarised ingredients for the prediction of the Higgs boson cross section at the LHC.
In order to derive such predictions it is key to asses all non-negligible sources of uncertainty. 
\ihixs{} allows to study such affects in great detail.
A careful analysis of residual uncertainties was performed in ref.~\cite{Anastasiou:2016cez} and we
implement the prescriptions chosen therein in our code. At the same time, \ihixs{} provides the user
with all tools neccessary to study individual sources of uncertainty and devise custom
prescriptions.
We briefly review the various sources of uncertainty in this section and describe our prescription to assess them quantitatively.

In ref.~\cite{Anastasiou:2016cez} the following sources of uncertainty were identified.
Missing higher order uncertainties are referred to as $\delta(\text{scale})$.
The uncertainty due to the evaluation of the Higgs boson production cross section at N$^3$LO with
PDFs determined with NNLO cross sections is denoted as $\delta(\text{PDF-TH})$.
$\delta(\text{EWK})$ indicates the uncertainty estimate for missing higher order mixed QCD and electro-weak corrections.
The quantity $\delta(\text{t,b,c})$ summarises the uncertainty due missing interference effects of top, bottom and charm quark masses at NNLO as well as the difference between different renormalisation schemes.
Missing effects due to the full top quark mass dependence of the Higgs boson cross section at NNLO
are estimated to introduce an uncertainty that we denote as $\delta(1/m_t)$.
The fully correlated combination of these sources of uncertainty form a quantity we refer to as theory uncertainty.
\beq
\delta(\text{theory})=\delta(\text{scale})+\delta(\text{PDF-TH})+\delta(\text{EWK})+\delta(\text{t,b,c})+\delta(1/m_t)\,.
\label{eq:TheoryUncertainty}
\eeq
Additional sources of uncertainty arise due to the imprecise knowledge of the strong couple constant $(\delta(\alpha_S))$ and the parton distribution functions $(\delta(\text{PDF}))$.
The fully uncorrelated combination of these uncertainties is given by
\beq
\delta(\text{PDF+$\alpha_S$})=\sqrt{\delta(\alpha_S)^2+\delta(\text{PDF})^2}\,.
\label{eq:AsPDFUncertainty}
\eeq

In combination we define the uncertainty estimate on the prediction for the inclusive production cross section for the Higgs boson.
\beq
\delta\sigma_{PP\rightarrow H+X}=\delta(\text{PDF+$\alpha_S$})+\delta(\text{theory})\,.
\label{eq:FullUncertainty}
\eeq
\ihixs{} provides the user with routines to estimate all of the above uncertainties.

Parametric uncertainties, due to imprecise knowledge of input parameters, are below one permille relative to the inclusive cross section for all reasonable scenarios in the Standard Model. 
Furthermore,
the size of the parametric uncertainty is mostly determined by the prior uncertainty estimate for
the input parameters. If the user would like to quantify the impact of the parametric uncertainty in
a particular prescription, this can be achieved straightforwardly, by simply varying the respective
input parmeters within their prior across multiple invocations of \ihixs{}.

\subsection{$\delta(\text{scale})$ - Missing Higher Orders}
\label{sec:ScaleUncertainty}
In \ihixs{} perturbative corrections in QCD can be included through N$^3$LO. 
Due to the trunctation of the perturbative series an uncertainty is introduced. 
In ref.~\cite{Anastasiou:2016cez} several options to estimate the effect of missing higher orders were explored. 
This analysis suggested that the size of the effect of missing higher orders can be estimated by
varying the common perturbative scale $\mu_F=\mu_R=\mu$ around the the central scale $\mu=\frac{m_h}{2}$.
We define
\bea
\sigma_{PP\rightarrow H+X}^{\text{max}}&=&\max\limits_{\mu\in[m_h/4,m_h]} \sigma_{PP\rightarrow H+X}(\mu,\mu)\,.\nonumber\\
\sigma_{PP\rightarrow H+X}^{\text{min}}&=&\min\limits_{\mu\in[m_h/4,m_h]} \sigma_{PP\rightarrow H+X}(\mu,\mu)\,.
\eea
The uncertainty that is associated with neglecting missing higher order contributions is then defined as 
\beq
\delta(\text{scale})=\begin{array}{c}\sigma_{PP\rightarrow H+X}^{\text{max}}- \sigma_{PP\rightarrow H+X}(\frac{m_h}{2},\frac{m_h}{2}) \\ \sigma_{PP\rightarrow H+X}(\frac{m_h}{2},\frac{m_h}{2})-\sigma_{PP\rightarrow H+X}^{\text{min}}\end{array}\,.
\eeq
Naturally, this prescription leads to asymmetric intervals for the uncertainty estimates. 

\subsection{$\delta(\text{PDF-TH})$ - PDF Theory Uncertainties}
\label{sec:PDF-TH}
Currently, parton distribution functions are determined by comparing cross section predictions at NNLO to physical measurements.
Since \ihixs{} can derive predictions at N$^3$LO another source of uncertainty is introduced due to the missmatch to the order of the PDFs.
In order to estimate this particular uncertainty we can analyse what would have happened at one order less in the same situation.
To this end we determine the cross section through NNLO, $ \sigma_{PP\rightarrow H+X}^{(2),\,\,\text{EFT}}$, evaluated once with NNLO PDFs and once with NLO PDFs.
The difference of these two predictions serves as our estimator of this particular uncertainty.
\beq
\delta(\text{PDF-TH})=\pm \frac{1}{2} \left| \sigma_{PP\rightarrow H+X}^{(2),\,\,\text{EFT},\,\,\text{NNLO}}-\sigma_{PP\rightarrow H+X}^{(2),\,\,\text{EFT},\,\,\text{NLO}} \right|\,.
\eeq
The factor of $\frac{1}{2}$ serves as a suppression factor as we expect this effect to be reduced at N$^3$LO relative to NNLO.
Since N$^3$LO predictions are only available in the EFT we estimate this effect based on predictions using EFT partonic cross sections only.

\subsection{$\delta(\text{EWK})$ - Missing Higher Order Electro-Weak Effects}
In ref.~\cite{Anastasiou:2016cez} several options to asses the uncertainty due to missing higher order electro-weak effects were discussed.
As a result an uncertainty of one percent on the total cross section was assigned.
\beq
\label{eq:deltaEW}
\delta(\text{EWK})=\pm1\%\times \sigma_{PP\rightarrow H+X}\,.
\eeq

\subsection{$\delta(\text{t,b,c})$ - Light Quark Masses and Renormalisation Schemes}
In \ihixs{} the effects of light quark masses are included exactly through NLO in QCD. 
In order to derive an estimate for the size of contributions due to finite light quark masses at NNLO we study how big the relative impact of light quarks on the NLO correction is.
We then assume that the relative impact of the light quark masses on NNLO corrections would be equally large and use this as an estimate of uncertainty.
\beq
\delta(t,b,c)^{\overline{\text{MS}}}=\pm\left|\frac{\delta \sigma^{t,\,\,\text{NLO}}-\delta\sigma^{t,b,c,\,\,\text{NLO}}}{\delta \sigma^{t,\,\,\text{NLO}}}\right|\times \left(\text{R}_{\text{LO}}\delta \sigma^{EFT,\,\,\text{NNLO}}+\delta \sigma^{1/m_t^2,\,\,\text{NNLO}}\right)\,.
\eeq
Here, $\delta \sigma^{t,\,\,\text{NLO}}$ and $\delta \sigma^{t,b,c,\,\,\text{NLO}}$ are the NLO  QCD corrections to the hadronic cross section with finite top quark mass and with finite top, bottom and charm quark mass respectively.
The corrections $\delta \sigma^{EFT,\,\,\text{NNLO}}$ refer to contributions to the hadronic cross section due to the EFT QCD corrections at NNLO.
Similarly, $\delta \sigma^{1/m_t^2,\,\,\text{NNLO}}$ describes QCD corrections at NNLO to the hadronic cross section due to the approximation of the exact NNLO cross section that are suppressed in powers of $1/m_t^2$.
To derive this estimate we work in the $\overline{\text{MS}}$ scheme.

Due to the truncation of the perturbative series a finite dependence on the chosen mass renormalisation scheme is introduced. 
To investigate the size of this dependence \ihixs{} includes implementations of quark mass effects in the $\overline{\text{MS}}$ and in the on-shell scheme.
In ref.~\cite{Anastasiou:2016cez} it was observed that choosing different renormalisation schemes for the top quark mass has negligible impact on the hadronic cross section.
In contrast, light quark mass effects display a more significant dependence on the choice of the renormalisation scheme.
The effects can reach up to thirty percent of these contributions.
In order to derive a conservative estimate of missing higher order contributions for light quark mass effects we multiply the above uncertainty by a factor of $1.3$
\beq
\label{eq:deltaTBC}
\delta(t,b,c)=1.3\times\delta(t,b,c)^{\overline{\text{MS}}}\,.
\eeq

\subsection{$\delta(1/m_t)$ - Missing Quark Mass Effects}
Effects due to the approximate treatement of QCD corrections at NNLO as an expansion in inverse powers of the top quark mass were studied in refs.~\cite{Harlander:2009my,Pak:2009dg}.
The consensus is that a residual uncertainty of one percent should be assigned to the cross section.
\beq
\label{eq:deltaMt}
\delta(1/m_t)=\pm1\%\times \sigma_{PP\rightarrow H+X}\,.
\eeq


\subsection{$\delta(\alpha_S)$ and $\delta(\text{PDF})$}
The estimation procedure of uncertainties due to the imprecise knowledge of parton distribution functions varies for different PDF sets.
\ihixs{} offers the possibility to automatically estimate the associated uncertainty $\delta(\text{PDF})$ using the default LHAPDF routines if available.

Assessing the uncertainty due to the prior uncertainty on the strong coupling constant typically requires a consistent treatment of parton distribution functions along with variations of $\alpha_S$.
The various groups providing fitted parton distribution functions recommend different procedures. 
In \ihixs{} we include routines that allow to automatically derive an uncertainty due to the uncertain strong coupling input value for the PDF4LHC15 PDF set~\cite{Butterworth:2015oua} at NNLO. 
A central value of $\alpha_S(m_Z)=0.118$ is chosen and the cross section predictions are also carried out with variations of the strong coupling constant by $\pm 0.0015$ including the usage of two dedicated PDF sets.
The strong coupling constant uncertainty is then given by
\beq
\delta(\alpha_S)=\frac{1}{2}\left|\sigma_{PP\rightarrow H+X}(\alpha_S(m_Z)=0.1195)-\sigma_{PP\rightarrow H+X}(\alpha_S(m_Z)=0.1165)\right|.
\eeq
If any other PDF set is chosen this uncertainty is not estimated automatically and the user has to follow their own procedure.

\section{Predictions for the LHC}
\label{sec:pheno}
In the previous sections we listed the various ingredients included in \ihixs{}. 
Here, we utilise our program to demonstrate the output that can be generated and derive state of the art predictions for the inclusive production probability of a Higgs boson at the LHC due to the gluon-fusion production mechanism.

Throughout this section we use PDF4LHC15 parton distribution functions~\cite{Butterworth:2015oua} at NNLO. We choose a value of the strong coupling constant of $\alpha_S(m_Z)=0.118$ and a Higgs boson mass of $m_h= 125$  GeV. 
The non-vanishing quark masses need to be specified at a reference scale $Q_0$. We use the values
given in table~\ref{tab:masses} in the $\overline{\textrm{MS}}$ scheme~\cite{Denner:2047636,deFlorian:2016spz}.
\begin{table}
    \begin{center}
        \begin{tabular}{lcc}
            \toprule
        &$m_q(Q_0)/\textrm{GeV}$ & $Q_0/\textrm{GeV}$\\
        \midrule
        $t$ & 162.7 & 162.7\\
        $b$ & 4.18 & 4.18\\
        $c$ & 0.986 & 3.0\\
        \bottomrule
    \end{tabular}
    \caption{Default values for the quark masses and starting scales for the respective evolutions of the masses.}
\label{tab:masses}
\end{center}
\end{table}
To derive cross section predictions we choose $\mu_R=\mu_F=m_h/2$ as central scales.

\begin{figure}
    \includegraphics[width=1.0\textwidth]{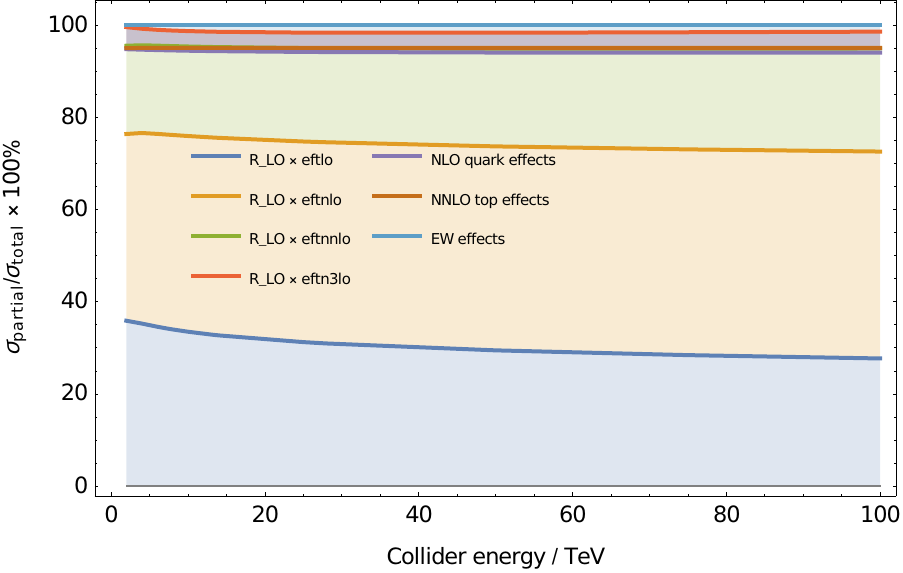}[h!]
    \caption{Relative cummulative contributions to the total cross section as a function of the collider
        energy.
    \label{fig:stack}}
\end{figure}

With a single run of \ihixs{} we can determine that the Higgs boson production cross section at the LHC with a center of mass energy of $13$ TeV is given by
\beq
\begin{array}{ccccc}
\sigma_{PP\rightarrow H+X}
&=& 16.00 \text{ pb} & (+ 32.87 \%) & \text{LO, rEFT}\\
&+& 20.84 \text{ pb} & (+ 42.82 \%) & \text{NLO, rEFT}\\
&+& 9.56 \text{ pb} & (+ 19.64 \%) & \text{NNLO, rEFT}\\
&+& 1.62 \text{ pb} & (+ 3.32 \%) & \text{N$^3$LO, rEFT}\\
&-& 2.07 \text{ pb} & (- 4.25 \%) & \text{(t,b,c) corr. to exact NLO}\\
&+& 0.34 \text{ pb} & (+ 0.70 \%) & \text{$1/m_t$ corr. to NNLO}\\
&+& 2.37 \text{ pb} & (+ 4.87 \%) & \text{EWK corr.}\\
&=& 48.67 \text{ pb}\,.& &
\end{array}
\eeq
Here effects from perturbative QCD through N$^3$LO, electro-weak interactions and finite quark
masses were taken into account as described in the previous sections. Figure~\ref{fig:stack} shows
the relative contributions of the the different components of the cross section as a function of the
collider energy; the data for such a plot is readily obtained by running \ihixs{} a few times for
different values of the collider energy.

\begin{figure}
    \includegraphics[width=1.0\textwidth]{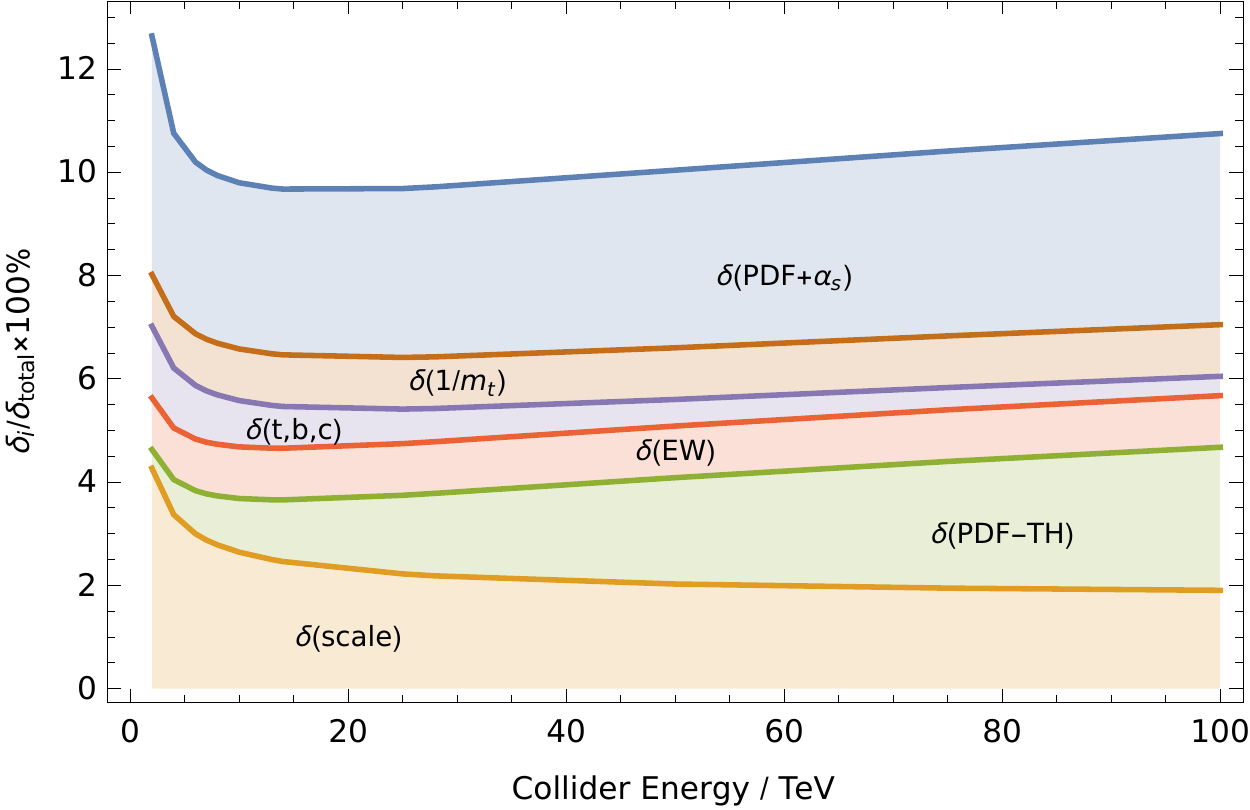}
    \caption{Cummulative contributions to the total relative uncertainty as a function of the collider energy.
        \label{fig:errorplot} according to eqs.~\eqref{eq:TheoryUncertainty}-\eqref{eq:FullUncertainty}.}
\end{figure}

From a single run of \ihixs{} we also obtain estimates for the residual uncertainty on the cross section. \ihixs{} provides detailed estimates for the various sources of uncertainty
\beq
\begin{array}{ccccc}
    \delta(\text{theory})&=&{}^{+0.13 pb}_{-1.20 pb} & \left({}^{+0.28 \%}_{-2.50 \%}\right) & \text{$\delta$(scale)} \\
    &+&\pm0.56pb &\left(\pm1.16\%\right)&\text{$\delta$(PDF-TH)}\\
    &+&\pm0.49pb &\left(\pm1.00\%\right)&\text{$\delta$(EWK)}\\
    &+&\pm0.41pb &\left(\pm0.85\%\right) &\text{$\delta$(t,b,c)}\\
    &+&\pm0.49pb &\left(\pm1.00\%\right) &\text{$\delta(1/m_t)$}\\
    &=&{}^{+2.08 pb }_{-3.16 pb} & \left( {}^{+4.28 \%}_{-6.5 \%}\right)\,,  &\\
    \delta(\text{PDF})&=&\pm0.89 \text{pb}&(\pm 1.85\%)\,,&\\
    \delta(\alpha_S)&=&{}^{+1.25pb}_{-1.26pb}&\left({}^{+2.59\%}_{-2.62\%}\right)\,.&\\
\end{array}
\eeq
In combination we find
\beq
\delta\sigma_{PP\rightarrow H+X}=\delta(\text{PDF+$\alpha_S$})+\delta(\text{theory})={}^{+3.63 pb }_{-4.72 pb}\,\, \left({}^{+7.46 \%}_{-9.7 \%}\right)\,.
\eeq
To derive the various sources of uncertainties we followed the prescriptions outlined above.
In fig.~\ref{fig:errorplot} we show how the relative size of the various sources of uncertainty varies as a function of the hadron collider energy.

In comparison to the numerical cross section predictions derived in ref.~\cite{Anastasiou:2016cez} we observe only minor changes. 
The difference arise solely due to the exact computation of the N$^3$LO QCD corrections in the heavy top quark effective theory obtained in ref.~\cite{Mistlberger:2018}. 
The deviations are well within the uncertainty that was associated with the truncation of the threshold expansion used for the results of ref.~\cite{Anastasiou:2016cez}. 
This particular source of uncertainty is now removed.
\begin{figure}[!t]
    \includegraphics[width=1.0\textwidth]{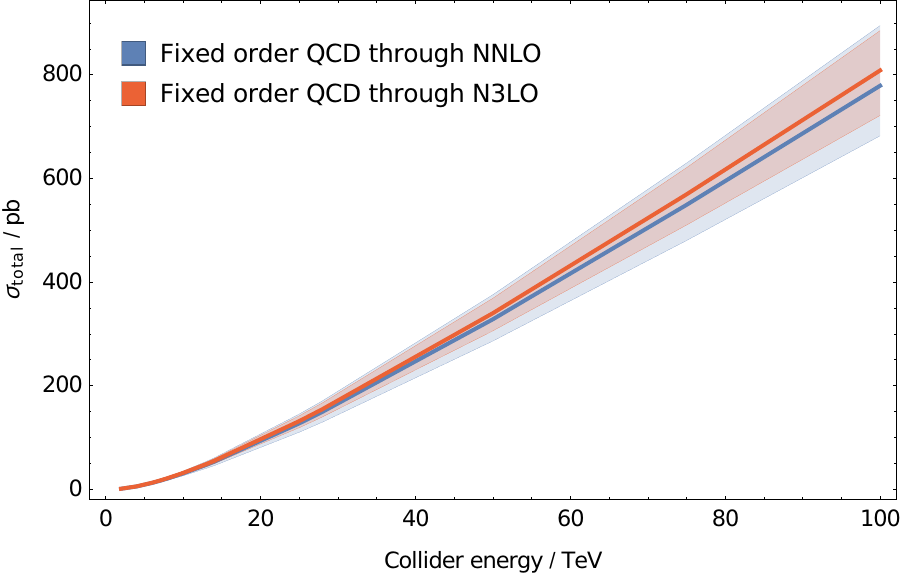}
    \caption{Total cross section and total at NNLO and N$^3$LO as a function of the collider energy. The bands show the combination of total uncertainty according to eq.~\eqref{eq:FullUncertainty}.
        \label{fig:eplot}}
\end{figure}

Finally, we use \ihixs{} to derive state of the art predictions for the gluon fusion Higgs production cross section at different collider energies. We strictly follow the recommendations of~\cite{Anastasiou:2016cez,deFlorian:2016spz}. Figure~\ref{fig:eplot} shows the state-of-the art predictions and uncertainty estimates for the inclusive cross section obtained this way compared to the prediction that is obtained without knowledge of the N$^3$LO corrections. In table~\ref{tab:results} we give the detailed numbers for the cross section and uncertainty estimates obtained with \ihixs{} including fixed order QCD corrections through N$^3$LO.
\begin{table}[!h]
\normalsize\setlength{\tabcolsep}{2pt}
\begin{center}
    \begin{tabular}{rrrrr}
        \toprule
        \multicolumn{1}{c}{$\textrm{E}_{\textrm{CM}}$}&
        \multicolumn{1}{c}{$\sigma$}&
        \multicolumn{1}{c}{$\delta(\textrm{theory})$}&
        \multicolumn{1}{c}{$\delta(\textrm{PDF})$}&
        \multicolumn{1}{c}{$\delta(\alpha_s)$}\\\midrule
        2 TeV & 1.10 pb & ${}^{+0.05\textrm{pb}}_{-0.09\textrm{pb}}\,\left({}^{+4.17\%}_{-8.02\%}\right) $ & $\pm\,0.03\,\textrm{pb}\,(\pm\,3.17\%)$ & ${}^{+0.04\textrm{pb}}_{-0.04\textrm{pb}}\,\left({}^{+3.69\%}_{-3.36\%}\right) $ \\\midrule
        7 TeV & 16.87 pb & ${}^{+0.70\textrm{pb}}_{-1.14\textrm{pb}}\,\left({}^{+4.17\%}_{-6.76\%}\right) $ & $\pm\,0.31\,\textrm{pb}\,(\pm\,1.89\%)$ & ${}^{+0.44\textrm{pb}}_{-0.45\textrm{pb}}\,\left({}^{+2.66\%}_{-2.68\%}\right) $ \\\midrule
8 TeV & 21.45 pb & ${}^{+0.90\textrm{pb}}_{-1.43\textrm{pb}}\,\left({}^{+4.18\%}_{-6.69\%}\right) $ & $\pm\,0.40\,\textrm{pb}\,(\pm\,1.87\%)$ & ${}^{+0.56\textrm{pb}}_{-0.56\textrm{pb}}\,\left({}^{+2.63\%}_{-2.66\%}\right) $ \\\midrule
13 TeV & 48.68 pb & ${}^{+2.07\textrm{pb}}_{-3.16\textrm{pb}}\,\left({}^{+4.26\%}_{-6.48\%}\right) $ & $\pm\,0.89\,\textrm{pb}\,(\pm\,1.85\%)$ & ${}^{+1.25\textrm{pb}}_{-1.26\textrm{pb}}\,\left({}^{+2.59\%}_{-2.62\%}\right) $ \\\midrule
14 TeV & 54.80 pb & ${}^{+2.34\textrm{pb}}_{-3.54\textrm{pb}}\,\left({}^{+4.28\%}_{-6.46\%}\right) $ & $\pm\,1.00\,\textrm{pb}\,(\pm\,1.86\%)$ & ${}^{+1.40\textrm{pb}}_{-1.42\textrm{pb}}\,\left({}^{+2.60\%}_{-2.62\%}\right) $ \\\midrule
28 TeV & 154.63 pb & ${}^{+7.02\textrm{pb}}_{-9.93\textrm{pb}}\,\left({}^{+4.54\%}_{-6.42\%}\right) $ & $\pm\,2.98\,\textrm{pb}\,(\pm\,1.96\%)$ & ${}^{+4.10\textrm{pb}}_{-4.03\textrm{pb}}\,\left({}^{+2.70\%}_{-2.65\%}\right) $ \\\midrule
100 TeV & 808.23 pb & ${}^{+44.53\textrm{pb}}_{-56.95\textrm{pb}}\,\left({}^{+5.51\%}_{-7.05\%}\right) $ & $\pm\,19.98\,\textrm{pb}\,(\pm\,2.51\%)$ & ${}^{+24.89\textrm{pb}}_{-21.71\textrm{pb}}\,\left({}^{+3.12\%}_{-2.72\%}\right) $ \\\bottomrule
    \end{tabular}
    \caption{Cross sections and uncertainties as function of the collider center of mass energy.\label{tab:results}}
\end{center}
\end{table}

\section{User-Manual for \ihixs }
\label{sec:manual}
\subsection{Prerequisites}
\ihixs{} is computing the hadronic inclusive Higgs cross-section, and therefore it depends on the
\texttt{LHAPDF6}  library~\cite{Buckley:2014ana} for initial state parton densities. Due to the
change in the interface of the \texttt{LHAPDF} library from version 5 to version 6, \ihixs{} is not
compatible with \texttt{LHAPDF} version $5.\ast$ or lower.   \ihixs{} also depends on the
\texttt{C++}  library \texttt{Boost} version 1.6 or higher\footnote{\url{http://www.boost.org/}}, which is also a dependency for \texttt{LHAPDF6}. 

Another requirement is a working version of the \texttt{CUBA} library of version 4.0 or higher for multidimensional integration~\cite{Hahn:2004fe}. In particular we use the implementations of the \texttt{Vegas} and \texttt{Cuhre}  algorithms provided by the library. 

Finally, the configuration step of the installation procedure is performed by \texttt{cmake}\footnote{\url{https://cmake.org/}}.

\subsection{Installation}
\ihixs{} can be obtained from \url{https://github.com/dulatf/ihixs}, either by downloading a gzipped release or cloning the repository with \texttt{git} using
\begin{lstlisting}
    git clone https://github.com/dulatf/ihixs    \end{lstlisting}

If all dependencies (i.e. \texttt{LHAPDF6} and \texttt{Cuba}) are globally installed, i.e. they can be found in the library path of the system, then configuration and building is performed by 
\begin{lstlisting}
    cd <PATH_TO_ihixs_SRC>
    mkdir build
    cd build
    cmake ..
    make
\end{lstlisting}
where \texttt{<PATH\_TO\_ihixs\_SRC>} is the path to the directory containing the \ihixs{} source
files (i.e. containing \texttt{ihixs.cpp}). An executable called \texttt{ihixs} is then generated at the current directory, along with various configuration cache files.  

If any of the dependencies are not installed globally, the paths to the respective libraries can be
specified in configuration time using the \texttt{-Dvar=value} flags of \texttt{cmake}. To
explicitly specify the location of all dependencies, \texttt{cmake} can be invoked as 
\begin{lstlisting}
    cmake -DLHAPDF_DIR=<lhapdf main dir> -DCUBA_DIR_USER=<cuba dir> -DBOOST_DIR_USER=<boost dir> <PATH_TO_ihixs_SRC> 
\end{lstlisting}
followed by 
\begin{lstlisting}
    make
\end{lstlisting}

A common source of installation problems in systems using the \texttt{Mac OS} comes from library
incompatibility between one of the dependencies built by the gnu compiler \texttt{gcc} and \ihixs{}
build by the native \texttt{LLVM} compiler of Apple or vice versa. This results in a linker error. Naturally the only way to fix the issue is to ensure that all components are build by the same (or at least compatible) compilers. One can force \texttt{cmake} to use a particular compiler by \emph{pre-pending} the paths to that compiler when invoking \texttt{cmake}:

\begin{lstlisting}
    CXX=<mygxx> CC=<mygcc> cmake  <PATH_TO_ihixs_SRC>
    make
\end{lstlisting}

Optionally, the user can include various tests in the build process, under the framework of \texttt{gtest}\footnote{https://github.com/google/googletest}. A version of the \texttt{gtest} unit test framework is distributed with the source code of \ihixs{}. To include the tests use 
\begin{lstlisting}
    cmake -Dwith_google_tests="true" <PATH_TO_ihixs_SRC>
    make
\end{lstlisting}
After compilation an extra set of executables is created in the directory \texttt{src/tests}. The user can verify then that the code produced passes all the tests by running them. For example 
 \begin{lstlisting}
    ./src/tests/ihixs_eft \end{lstlisting}
should run various tests related to the implementation of the coefficient functions necessary for the computation of the Higgs cross section within the EFT approximation. All tests should pass and the user should recive a message similar to 
  \begin{lstlisting}
[----------] Global test environment tear-down
[==========] 69 tests from 20 test cases ran. (166 ms total)
[  PASSED  ] 69 tests.\end{lstlisting}

We have checked that \ihixs{} compiles properly under \texttt{linux} with the generic gnu compiler \texttt{gcc 4.8} and higher, and under \texttt{Mac OS X} with both \texttt{gcc 4.8} and \texttt{LLVM} version \texttt{9.0} or higher.

\subsection{Usage}

Input parameters for \ihixs{} are specified in a runcard file, as well as by command line options. There is a default runcard called \texttt{default.card} supplied in the distribution\footnote{The user can always regenerate this default card by \texttt{ihixs --make\_runcard} or with the shorthand \texttt{ihixs -d}.}. 

The user can run \ihixs{}, using the default settings\footnote{Please note that the default settings do \textbf{NOT} reproduce the full Higgs inclusive cross-section, but only the rescaled EFT approximation of it. } by 
 \begin{lstlisting}
   ./ihixs -i default.card\end{lstlisting}
 In the absence of a \texttt{default.card} file, the program can be run with the same default settings by
 \begin{lstlisting}
   ./ihixs \end{lstlisting}
 After half a minute or so the user should get the output
 \begin{lstlisting}
Result
mur                       = 62.5
muf                       = 62.5
R_LO                      = 1.06274
R_LO*eftlo                = 15.9988 [0.00158078]
R_LO*eftnlo               = 36.8376 [0.00205626]
R_LO*eftnnlo              = 46.3981 [0.00479644]
R_LO*eftn3lo              = 48.0162 [0.0116302]
Higgs XS                  = 48.0162 [0.0116302]
-------------------------------------------------
Higgs_XS = 48.0162\end{lstlisting}
The main result is the Higgs boson production cross section: \texttt{Higgs XS}. Computation of the theory uncertainty is switched off in this default settings. 

The numbers in brackets that follow the results are numerical integration errors (propagated properly for each quantity) and can be reduced by increasing the precision of numerical integration using the numerical precision options. 

The full phenomenological prediction for the Higgs cross section at the LHC, at $13$TeV,  including all theory uncertainties, is computed by running with the \texttt{pheno.card} that is provided in the distribution:
 \begin{lstlisting}
     ./ihixs -i pheno.card\end{lstlisting}
 The output is now
 \begin{lstlisting}
Result
mur                       = 62.5
muf                       = 62.5
R_LO                      = 1.06274
R_LO*eftlo                = 15.9988 [0.00158078]
R_LO*eftnlo               = 36.8376 [0.00205626]
R_LO*eftnnlo              = 46.3981 [0.00479644]
R_LO*eftn3lo              = 48.0162 [0.0116302]
ew rescaled               = 2.37184 [0.000609145]
NLO quark mass effects    = -2.06583 [0.00509613]
NNLO top mass effects     = 0.343941 [0.00450368]
Higgs XS                  = 48.6661 [0.0134866]
delta_tbc                 = 0.411289 [0.00428913]
delta_tbc %               = 0.845123 [0.00881649]
delta(1/m_t)              = 0.486661 [0.000134866]
delta(1/m_t) %            = 1 [0]
delta EW                  = 0.486661 [0.000134866]
delta EW %                = 1 [0]
delta PDF-TH %            = 1.15673 [0.00738757]
delta(scale)+             = 0.133725
delta(scale)-             = -1.19938
delta(scale)+(%)          = 0.2785
delta(scale)-(%)          = -2.49786
deltaPDF+                 = 0.888988
deltaPDF-                 = 0.888988
deltaPDFsymm              = 0.888988
deltaPDF+(%)              = 1.85134
deltaPDF-(%)              = 1.85134
delta(as)+                = 1.24558
delta(as)-                = -1.25836
delta(as)+(%)             = 2.59407
delta(as)-(%)             = -2.62071
Theory Uncertainty  +     = 2.08308 [0.00562749]
Theory Uncertainty  -     = -3.16316 [0.00566602]
Theory Uncertainty % +    = 4.28035 [0.0115025]
Theory Uncertainty % -    = -6.49972 [0.0115025]
delta(PDF+a_s) +          = 1.55097 [0.00042981]
delta(PDF+a_s) -          = -1.56154 [-0.000432739]
delta(PDF+a_s) + %        = 3.18695 [0]
delta(PDF+a_s) - %        = -3.20867 [0]
Total Uncertainty +       = 3.63405 [0.00564388]
Total Uncertainty -       = -4.7247 [0.00568253]
Total Uncertainty + %     = 7.46731 [0.0115025]
Total Uncertainty - %     = -9.70839 [0.0115025]

-------------------------------------------------
Higgs_XS = 48.6661 +3.634(7.5%) -4.725(-9.7%)\end{lstlisting}
The  \texttt{Higgs XS} and the \texttt{Theory Uncertainty} outputs include all the effects specified by the operational options in the runcard. 
Note that the total cross section \texttt{Higgs XS} is the sum of all the contributions of eq.~\ref{eq:masterdec}. Moreover the uncertainties are computed as described in section~\ref{sec:Uncertainties} and combined according to eq.~\ref{eq:TheoryUncertainty},~\ref{eq:AsPDFUncertainty} and~\ref{eq:FullUncertainty}.
 
In case of undesired modification the \texttt{pheno.card} can be recreated with 
  \begin{lstlisting}
    ./ihixs --make_pheno.card\end{lstlisting}
 or the shorthand version
   \begin{lstlisting}
    ./ihixs -p\end{lstlisting}
See section~\ref{sec:ioandvars} for details on command line options.


\subsubsection{Input options and variables}
\label{sec:ioandvars}
There are four classes of input options: operational options, input-output options, masses-and-scales options and numerical precision options. All available options with some explanation about their functionality can be found in Tables~\ref{tab:operationaloptions},~\ref{tab:inputoutputoptions},~\ref{tab:massesandscalesoptions} and~\ref{tab:numericalprecisionoptions}. They can also be inspected by 
 \begin{lstlisting}
    ./ihixs --help \end{lstlisting}
 
All options can also be set at the command line, and if so, the value provided overwrites the one in the runcard (irrespectively of the order of command line arguments). For example one can run specifying that the output file should be \texttt{new\_out.txt} by 
  \begin{lstlisting}
    ./ihixs --output_filename new_out.txt --input_filename default.card \end{lstlisting}
 
 Some options that are frequently used have shorthands. These shorthands appear in square brackets at the output of \texttt{ihixs --help} and  they should be used with a single `\texttt{-}'. So for example one can specify the output file using the shorthand 
   \begin{lstlisting}
   ./ihixs -o new_out.txt -i default.card\end{lstlisting}

\subsubsection{Options related to the total theory uncertainty}
Theory uncertainties are switched on by default if the corresponding contributions that give rise to these uncertainties are included in the computation: if the computation includes the quark mass effects at NLO, then the uncertainty due to the unknown NNLO mass effects is switched on (computed according to eq.~\ref{eq:deltaTBC}), if electroweak corrections are switched on, so is the uncertainty due to higher order, mixed QCD-EW corrections (computed according to eq.~\ref{eq:deltaEW}), and the same is true for top mass effects at NNLO (whose uncertainty is set ad hoc to $1\%$, see eq.~\ref{eq:deltaMt}). The remaining uncertainties, due to the lack of N3LO parton density fits, the perturbative scale, the fits of the existing parton densities  and the value of the strong coupling constant, have to be explicitly switched on, by setting to \texttt{true} the corresponding parameters: 

\begin{description}
    \item[\texttt{with\_delta\_pdf\_th}] Activates the computation of the PDF\_TH uncertainty as explained in section~\ref{sec:PDF-TH}. The PDF set used at NLO is the one specified by the input parameter \texttt{pdf\_set\_for\_nlo}. Note that no checks are performed regarding the compatibility of this PDF set with the one used for the main run, specified by \texttt{pdf\_set}. The user is, thus, responsible to select a compatible, nlo-level, PDF set.
    \item [\texttt{with\_scale\_variation}] Activates the computation of scale uncertainty. In order
        not to make the run unnecessarily slow, we compute the scale uncertainty by evaluating the
        rescaled EFT cross section at $\mu=0.25 m_H$, $\mu=0.4m_H$ and $\mu=m_H$. With reasonable sets of input
        variables the cross-section takes its maximal value around $\mu=0.4 m_H$ if N$^3$LO QCD corrections are included. 
        Consequently, this procedure guarantees that the scale uncertainty is very close, numerically, to that prescribed in
        section~\ref{sec:ScaleUncertainty}. At lower orders the extremal values of the cross section are found at the boundaries of the scale variation interval.
        If one does a scan over the scale $\mu$ it is possible
        to find slight differences (at the per mille level) with respect to the upper end of the
        uncertainty interval, because it is conceivable that the cross-section does not reach its
        maxiumum for $\mu=0.4 m_H$. Given that the dependence of the N$^3$LO cross section on the scale
        $\mu$ is so mild, this effect is most of the times entirely negligible. Still, if the user
        wants to experiment with wildly different parameters than those corresponding to the LHC set
        up, they should run a proper scale scan to estimate the scale uncertainty. The deviation from the central value of the rescaled EFT cross section for both the upper and the lower end of the (very asymmetric) uncertainty band, in absolute and relative terms is quoted in the output:
     \begin{lstlisting}
delta(scale)+        = 0.133725
delta(scale)-        = -1.19938
delta(scale)+(%)     = 0.2785
delta(scale)-(%)     = -2.49786
\end{lstlisting}   
Note that we do not consider an independent variation of renormalization and factorization scales, a choice that is justified by the very small dependence of the cross section to the factorization scale. The user can, of course, determine their own version of scale uncertainty by explicitly setting the values for the factorization scale, \texttt{muf} and the renormalization scale \texttt{mur}, in the runcard or through the command line parameters in multiple runs. 
    \item[\texttt{with\_pdf\_error}] Activates the computation of the PDF error. This assumes that the
        PDF set selected, through \texttt{pdf\_set} has more than one pdf member. The computation of
        the PDF uncertainty is delegated to the LHAPDF library, in order to use the procedure
        appropriate to the selected PDF set. The observable computed using the different PDF members is the rescaled EFT cross section to N3LO. 
The type of uncertainty computed, specified by LHAPDF, is outputed on screen at runtime. The user sees a message like
    \begin{lstlisting}
    Computing PDF error with PDF4LHC15_nnlo_100 which has 101 members
Description : PDF4LHC15_nnlo_100. mem=0 => alphas(MZ)=0.118 central value; mem=1-100 => PDF symmetric eigenvectors
Data version : 1
Error type : symmhessian    
\end{lstlisting}
The deviation from the central value of the rescaled EFT cross section for both the upper and the lower end of the uncertainty band, in absolute and relative terms is quoted in the output:
 \begin{lstlisting}
deltaPDF+            = 0.888988
deltaPDF-            = 0.888988
deltaPDFsymm         = 0.888988
deltaPDF+(%)         = 1.85134
deltaPDF-(%)         = 1.85134\end{lstlisting}
	\item[\texttt{with\_a\_s\_error}] Activates the computation of the uncertainty due to $a_s$. This assumes that the
        PDF set selected, through \texttt{pdf\_set} is the \\ \texttt{PDF4LHC15\_nnlo\_100\_pdfas} set. If \texttt{PDF4LHC15\_nnlo\_100} is declared instead, it is switched to \texttt{PDF4LHC15\_nnlo\_100\_pdfas} for the purposes of estimating the $a_s$ uncertainty. 
        
        The rescaled EFT cross section is computed with members $101$ and $102$ of this PDF set, that correspond to fits with the value of $a_s$ at the upper and lower bound of the variation range adopted by the \texttt{PDF4LHC} working group. The resulting rescaled EFT cross sections, as well as the deviations from the central value of the rescaled EFT cross section  in absolute and relative terms for both the upper and the lower end of the uncertainty band are quoted in the output: 
        \begin{lstlisting}
rEFT(as+)            = 49.2618 [0.0123911]
rEFT(as-)            = 46.7578 [0.0108436]
delta(as)+           = 1.24558
delta(as)-           = -1.25836
delta(as)+(%)        = 2.59407
delta(as)-(%)        = -2.62071	\end{lstlisting}
\end{description}

\subsubsection{Option related to quark mass effects}
\begin{description} \item[\texttt{with\_indiv\_mass\_effects}] Computes the exact NLO cross section in the presence of massive quarks, in all possible combinations. The results are reported as 
 \begin{lstlisting}
exact LO t           = 15.9988 [0.00158078]
exact NLO t          = 36.6011 [0.00452547]
delta QCD            = -2.06583 [0.00509613]
exact LO t+b         = 14.9428 [0.00147644]
exact NLO t+b        = 34.9627 [0.00464004]
exact LO t+c         = 15.8762 [0.00156866]
exact NLO t+c        = 36.3833 [0.0045734]
exact LO b           = 0.0423262 [4.18209e-06]
exact NLO b          = 0.101262 [9.72963e-06]
exact LO c           = 0.000347049 [3.42906e-08]
exact NLO c          = 0.000895551 [1.23147e-07]
exact LO b+c         = 0.0501185 [4.95201e-06]
exact NLO b+c        = 0.120868 [1.1858e-05]
\end{lstlisting}
For example \texttt{exact NLO b+c} is the cross section with the bottom and charm quarks activated (including their interference terms), while the top quark is de-activated. Note that setting the mass of any quark to exactly \texttt{0.0} leads to an program error. However, using any small, positive value for the quark masses is allowed and leads to valid results. Hence one can decouple the bottom quark, for example, by setting its mass to \texttt{0.001} (all masses are in GeV). On the other hand, it is more transparent to do so by setting the corresponding Yukawa rescaling coefficient, \texttt{y\_bot} to \texttt{0.0}. 
\item \texttt{mt\_msbar, mt\_msbar\_ref\_scale} This is an input parameter for the $\overline{\text{MS}}$ mass of the top quark and the reference scale at which this mass is defined. Similar parameters exist for the bottom and the charm quarks. The program automatically evolves the quark mass from the reference scale to the renormalization scale, defined by \texttt{mur}. The default values for the masses and the reference scales are described in table \ref{tab:masses}. 

\item \texttt{mt\_on\_shell} This is the on-shell mass for the top quark. Similar parameters exist for the bottom and charm quarks. The default on-shell mass for the quarks are selected so that they are compatible with the default $\overline{\text{MS}}$ masses. 

\item \texttt{top\_scheme} This determines whether the on-shell scheme or the $\overline{\text{MS}}$ scheme will be used throughout the computation. Quantities that are scheme dependent, like the NLO quark mass effects, are consistently computed depending on the choice set by this option. 

\item \texttt{gamma\_top} The value of the top width. Similar parameters exist for the bottom and charm quarks. This is included in the LO and NLO computations by turning the quark mass into a complex parameter. The effect of non-zero width for all three quarks is below the sub-per-mille level, so the widths for all quarks are set to zero by default, for efficiency. 
\end{description}

\subsubsection{Results and output}
Apart from the output in the standard terminal, \ihixs{} also writes output in a more detailed format, at an output file whose name is determined by the \texttt{output\_filename} option. Note that the program will overwrite any existing file with the same name. The default output filename is \texttt{ihixs\_output}. 

The output file consists of three sections. The first section is a more detailed, human-readable,
version of the output that is written on screen during the calculation:
\begin{lstlisting}
ihixs results 
Result
mh                        = 125
Etot                      = 13000
PDF set                   = PDF4LHC15_nnlo_100
PDF member                = 0
mur                       = 62.5
muf                       = 62.5
as_at_mz                  = 0.118002
as_at_mur                 = 0.125161
mt_used                   = 176.416
mb_used                   = 2.96088
mc_used                   = 0.654089
eftlo                     = 15.0543 [0.00148746]
eftnlo                    = 34.6629 [0.00193487]
eftnnlo                   = 43.659 [0.00451328]
eftn3lo                   = 45.1816 [0.0109436]
R_LO                      = 1.06274
R_LO*eftlo                = 15.9988 [0.00158078]
R_LO*eftnlo               = 36.8376 [0.00205626]
R_LO*eftnnlo              = 46.3981 [0.00479644]
R_LO*eftn3lo              = 48.0162 [0.0116302]
ggnlo/eftnlo              = 0.959267 [7.70006e-05]
qgnlo/eftnlo              = 0.0400834 [4.56688e-06]
ggnnlo/eftn2lo            = 0.947457 [0.000142008]
qgnnlo/eftn2lo            = 0.0498868 [6.66652e-06]
ggn3lo/eftn3lo            = 0.947899 [0.000255586]
qgn3lo/eftn3lo            = 0.0482315 [1.26758e-05]
R_LO*gg channel           = 45.5145 [0.00539197]
R_LO*qg channel           = 2.31589 [0.000236196]
R_LO*qqbar channel        = 0.0464633 [0.000468206]
R_LO*qq channel           = 0.0353943 [0.000268044]
R_LO*q1q2 channel         = 0.103941 [0.010288]
ew rescaled as^2          = 0.83795 [8.27946e-05]
ew rescaled as^3          = 1.03791 [6.88804e-05]
ew rescaled as^4          = 0.436607 [0.00022696]
ew rescaled as^5          = 0.0593688 [0.000554931]
mixed EW-QCD              = 1.53389 [0.000614746]
ew rescaled               = 2.37184 [0.000609145]
hard ratio from eft       = 0.452463 [4.91477e-05]
WC                        = 1.11002 [0]
WC^2                      = 1.23216 [0]
WC^2_trunc                = 1.23213 [0]
n                         = 36.7594 [0.0109214]
sigma factorized          = 45.2932 [0.0111813]
exact LO t+b+c            = 14.8276 [0.00146506]
exact NLO t+b+c           = 34.7718 [0.00466286]
exact LO t                = 15.9988 [0.00158078]
exact NLO t               = 36.6011 [0.00452547]
NLO quark mass effects    = -2.06583 [0.00509613]
NLO quark mass effects / eft % = -5.60793
delta sigma t NLO         = 20.6023 [0.00479362]
delta sigma t+b+c NLO     = 19.9442 [0.00488761]
delta tbc ratio           = 0.031943 [0.000332376]
NNLO mt exp gg            = 0.387362 [0.0045011]
NNLO mt exp qg            = -0.0434209 [0.000152313]
NNLO top mass effects     = 0.343941 [0.00450368]
Higgs XS                  = 48.6661 [0.0134866]
delta_tbc                 = 0.411289 [0.00428913]
delta_tbc %               = 0.845123 [0.00881649]
delta(1/m_t)              = 0.486661 [0.000134866]
delta(1/m_t) %            = 1 [0]
delta EW                  = 0.486661 [0.000134866]
delta EW %                = 1 [0]
R_LO*eftnnlo (with NLO PDF) = 47.4715 [0.00489674]
delta PDF-TH %            = 1.15673 [0.00738757]
rEFT(low)                 = 48.1499 [0.0117723]
rEFT(high)                = 46.8168 [0.0175512]
delta(scale)+             = 0.133725
delta(scale)-             = -1.19938
delta(scale)+(%)          = 0.2785
delta(scale)-(%)          = -2.49786
delta(scale)+ pure eft    = 0.243273
delta(scale)- pure eft    = -1.48381
delta(scale)+(%) pure eft = 0.538434
delta(scale)-(%) pure eft = -3.2841
R_LO*eftn3lo_central      = 48.0187 [0.00507515]
deltaPDF+                 = 0.888988
deltaPDF-                 = 0.888988
deltaPDFsymm              = 0.888988
deltaPDF+(%)              = 1.85134
deltaPDF-(%)              = 1.85134
rEFT(as+)                 = 49.2618 [0.0123911]
rEFT(as-)                 = 46.7578 [0.0108436]
delta(as)+                = 1.24558
delta(as)-                = -1.25836
delta(as)+(%)             = 2.59407
delta(as)-(%)             = -2.62071
Theory Uncertainty  +     = 2.08308 [0.00562749]
Theory Uncertainty  -     = -3.16316 [0.00566602]
Theory Uncertainty % +    = 4.28035 [0.0115025]
Theory Uncertainty % -    = -6.49972 [0.0115025]
delta(PDF+a_s) +          = 1.55097 [0.00042981]
delta(PDF+a_s) -          = -1.56154 [-0.000432739]
delta(PDF+a_s) + %        = 3.18695 [0]
delta(PDF+a_s) - %        = -3.20867 [0]
Total Uncertainty +       = 3.63405 [0.00564388]
Total Uncertainty -       = -4.7247 [0.00568253]
Total Uncertainty + %     = 7.46731 [0.0115025]
Total Uncertainty - %     = -9.70839 [0.0115025]
\end{lstlisting}
The first nine lines list information about the parameters used in this run and should be
self-explanatory. Afterwards, \ihixs{} lists the results of the current run.
The various values are as follows:
\begin{description}
	\item[ \texttt{as\_at\_mz}] The value of $a_s(m_Z)$, as retrieved from \texttt{LHAPDF}.
	\item[ \texttt{as\_at\_mur}] The value of $a_s(\mu_r)$, computed via evolution through N$^3$LO from $a_s(m_Z)$ quoted one line above. 
	\item[ \texttt{mt\_used, mb\_used, mc\_used}] The value of the quark mass, $m_q$, used in the computation. Depending on the scheme specified for each quark, this might be the on-shell mass (specified in the runcard) or the $\overline{\text{MS}}$ mass, evolved from its reference value $m_q(\mu_q)$ to $\mu_r$. Both $m_q$ and $\mu_q$ are also specified in the runcard (default values are recorded at Tab.~\ref{tab:masses}).
	    \item[\texttt{R\_LO}] The ratio of the exact LO cross section to the effect theory LO cross
        section.
    \item[\texttt{eftlo, eftnlo, eftnnlo, eftn3lo}] The cross section in the pure effective theory (no
        rescaling of the leading order) through N$^3$LO.
    \item[\texttt{R\_LO*eftlo},\ldots, \texttt{R\_LO*eftn3lo}] The cross section in the effective
        theory, multiplied by the ratio of the exact LO cross section to the effect theory LO cross
        section, through N$^3$LO.
    \item[\texttt{ggnlo/eftnlo}] The fraction of the EFT cross-section due to the gluon-gluon channel at NLO.
    \item[\texttt{qgnlo/eftnlo}] The fraction of the EFT cross-section due to the quark-gluon channel at NLO.
    \item[\texttt{ggnnlo/eftn2lo}] The fraction of the EFT cross-section due to the gluon-gluon channel at NNLO.
    \item[\texttt{qgnnlo/eftn2lo}] The fraction of the EFT cross-section due to the quark-gluon channel at NNLO.
    \item[\texttt{ggn3lo/eftn3lo}] The fraction of the EFT cross-section due to the gluon-gluon channel at N3LO.
    \item[\texttt{qgn3lo/eftn3lo}] The fraction of the EFT cross-section due to the quark-gluon channel at N3LO.
    \item[\texttt{R\_LO*gg channel}] The EFT contribution of the gluon-gluon channel rescaled by \texttt{R\_LO}.
    \item[\texttt{WC}] The Wilson coefficient at the current scale.
    \item[\texttt{WC\^{}2}] The square of the Wilson coefficient, not truncated.
    \item[\texttt{WC\^{}2\_trunc}] The square of the Wilson coefficient truncate to order $\mathcal{O}(a_s^6)$.
    \item[\texttt{sigma factorized}] The EFT cross-section computed in a factorized form\footnote{The squared Wilson coefficient, $C^2$ and the parton-level matrix elements $\eta_{ij}$ are computed as an expansion in $a_s$. Normally the two expansions are combined and the result is truncated at the desired order. By `factorized' here we mean that the cross-section is obtained by multiplying $C^2 \cdot \eta$, without truncating, so it differs from the \texttt{eftn3lo} above by terms of order higher than $a_s^5$}, i.e. as a product of $C^2 \cdot \eta$.
     \item[\texttt{n}] The numerical value of the parton-level matrix elements, $\eta$, i.e. the EFT cross section with the Wilson coefficient set to 1. 
    \item[\texttt{Higgs XS}] The total Higgs cross section including all effects that were switched on in the current run\footnote{\emph{Including} resummation effects if activated, either following classical resummation, or SCET type resummation.}.
   	\item[\texttt{exact LO t+b+c}]    The exact LO cross section including top and light quark mass effects.
	\item[\texttt{exact NLO t+b+c}]      The exact NLO cross section including top and light quark mass effects.
	\item[\texttt{exact LO t   }]             The exact LO cross section including top mass effects only.
	\item[\texttt{exact NLO t   }]             The exact NLO cross section including top mass effects only.
	\item[\texttt{NLO quark mass effects}]    Corrections to the rescaled EFT cross section due to mass effects, at NLO.
	\item[\texttt{NLO quark mass effects / eft \%}] \% effect from light quark masses  over the EFT cross section.
	\item[\texttt{NNLO mt exp gg}]            NNLO top mass effects in the gluon-gluon channel.
	\item[\texttt{NNLO mt exp qg  }]          NNLO top mass effects in the quark-gluon channel.
	\item[\texttt{NNLO top mass effects}]     NNLO top mass effects.
	\item[\texttt{Higgs XS}]                  Total Higgs cross section including all effects specified in the runcard.
	\item[\texttt{delta\_tbc, delta\_tbc \%}]                Uncertainty due to light quark mass effects.
    \item[\texttt{delta(1/m\_t), delta(1/m\_t) \%}]            Uncertainty due to top quark effects beyond NNLO.
	\item[\texttt{delta EW, delta EW \%}]               Uncertainty due to EW effects.
	\item[\texttt{R\_LO*eftnnlo (with NLO PDF)}]  NNLO EFT cross section rescaled by $R_{LO}$, computed with NLO PDFs, used in the computation of the PDF-TH error.
	\item[\texttt{delta PDF-TH \%}]            Uncertainty due to missing N$^3$LO PDFs.
    \item[\texttt{rEFT(low), rEFT(high) }]   minimum and maximum EFT cross section rescaled by $R_{LO}$, computed within the renormalization/factorization scale interval.
	\item[\texttt{delta(scale) }]    Scale uncertainty (rescaled by $R_{LO}$ ) in absolute and relative terms.
	\item[\texttt{delta(scale)+ pure eft }] Scale uncertainty for the EFT cross section (not rescaled by $R_{LO}$).
	\item[\texttt{deltaPDF}]   PDF uncertainty.
    \item[\texttt{rEFT(as+), rEFT(as-) }]   EFT cross section (rescaled by $R_{LO}$) computed with $a_s$ values at the edges of the $a_s$uncertainty interval. 
	\item[\texttt{delta(as)+ }]   Uncertainty due to $a_s$ in relative and absolute terms.
	\item[\texttt{Theory Uncertainty }] Total theory uncertainty computed as explained in eq.~\ref{eq:TheoryUncertainty}.
	\item[\texttt{delta(PDF+a\_s)  }] Combined PDF and $a_s$ uncertainty, computed as explained in eq.~\ref{eq:AsPDFUncertainty}.
	\item[\texttt{Total Uncertainty }] Total uncertainty computed as in eq.~\ref{eq:FullUncertainty}.
\end{description}

More information is displayed depending on the options activated in the current run. 

This first section of output in the output file is followed by the same output as above, but written in Mathematica format, using the dictionary data structure of Mathematica. This is very helpful when \ihixs{} is run on a cluster with many different input cards, as one would typically do to perform scans over one or more parameters. For example, if one wants to  get a detailed scan over the renormalization scale, one could run in a cluster with as many cards as there are points in the scan. It is easy to write a script to collect such results and import them to Mathematica for further analysis and plotting, by means of using the dictionary data structure provided. 

The next section of the output file is a list of the input parameters as they are used in the run. These might differ from parameters as they appear in the runcard, because upon running \ihixs{} they might have been overwritten by command-line parameters. Again, this typically happens when one runs on clusters to perform a parameter scan, and instead of running with many different runcards, one choses to run with the same runcard but modifying at command line the value of some parameter. For example, to perform a scan over different values for the renormalisation scale $\mu_R$ in conjunction with a runcard one could use the command
  \begin{lstlisting}
    ./ihixs -i my_generic_card -o results_for_mur_eq_<x> --mur=<x>  \end{lstlisting}
where a script is used to loop over different values of  the string \texttt{<x>}.

The filename of the runcard used in the run can be found at the bottom of the third section of the output file.

\begin{table}[h!]
\bgfb
\multicolumn{2}{|c|}{\textbf{Table~\ref{tab:operationaloptions}: Operational options}}
\\ 
\tt{Etot} : \textbf{13000.0} & COM energy of the collider in GeV
\\
\tt{pdf\_member} : \textbf{0} & pdf member id (the range depends on the pdf set)
\\
\tt{pdf\_set} : \textbf{PDF4LHC15\_nnlo\_100} & choose a specific pdf set name (LHAPDF6 list at lhapdf.hepforge.org/pdfsets.html). This set will be used irrespectively of order.
\\
\tt{pdf\_set\_for\_nlo} : \textbf{PDF4LHC15\_nlo\_100} & pdf set used when computing PDF-TH error.
\\
\tt{with\_eft} : \textbf{true} & compute the cross section in the EFT approximation
\\
\tt{with\_exact\_qcd\_corrections} : \textbf{false} & true to include the exact quark mass effects at NLO, false to omit them
\\
\tt{with\_ew\_corrections} : \textbf{false} & true to include the exact quark mass effects at NLO, false to omit them
\\
\tt{with\_mt\_expansion} : \textbf{false} & include NNLO 1/mt terms
\\
\tt{with\_delta\_pdf\_th} : \textbf{false} & compute PDF-TH uncertainty
\\
\tt{with\_scale\_variation} : \textbf{false} & estimate scale variation (mur and muf should be at mh/2)
\\
\tt{with\_indiv\_mass\_effects} : \textbf{false} & compute separately light quark contributions
\\
\tt{with\_pdf\_error} : \textbf{false} & whether or not to compute error due to pdfs
\\
\tt{with\_a\_s\_error} : \textbf{false} & compute a\_s uncertainty
\\
\tt{with\_resummation} : \textbf{false} & include threshold resummation
\\
\tt{resummation\_log\_order} : \textbf{3} & 0:LL, 1:NLL, 2:NNLL, 3:N3LL
\\
\tt{resummation\_matching\_order} : \textbf{3} & 0:L0, 1:NL0, 2:NNL0, 3:N3L0
\\
\tt{resummation\_type} : \textbf{log} & variant of threshold resummation, i.e. log:classical, psi, AP2log, AP2psi 
\\
\tt{with\_scet} : \textbf{false} & include scet resummation
\\
\tt{qcd\_perturbative\_order} : \textbf{N3LO} & LO, NLO, NNLO, N3LO : ihixs will compute up to this order in a\_s
\\
\tt{with\_fixed\_as\_at\_mz} : \textbf{0.0} & set the value of a\_s(mZ) by hand. Beware: this might not be compatible with your pdf choice.
\\
\tt{qcd\_order\_evol} : \textbf{3} & used for a\_s and quark mass evolution 0:L0, 1:NL0, 2:NNL0, 3:N3L0
\\
\tt{with\_lower\_ord\_scale\_var} : \textbf{false} & also compute scale variation for lower than the current order
\egfb
\fakecaption\label{tab:operationaloptions}
\end{table}
\begin{table}
\bgfb
\multicolumn{2}{|c|}{\textbf{Table~\ref{tab:inputoutputoptions}: Input-Output options}}
\\ 
\tt{verbose} : \textbf{minimal} & level of verbosity: minimal or medium. Medium shows channel breakdown EFT cross section.
\\
\tt{input\_filename} : \textbf{default.card} & filename to use as runcard
\\
\tt{output\_filename} : \textbf{ihixs\_output} & filename to write output
\\
\tt{help} : \textbf{false} & print all options and help messages per option.
\\
\tt{make\_runcard} : \textbf{false} & create default runcard file as default\_card.
\\
\tt{make\_pheno\_card} : \textbf{false} & create pheno runcard file as pheno\_card.
\\
\tt{write\_documentation} : \textbf{false} & print the help message in a TeX form.
\\
\tt{with\_eft\_channel\_info} : \textbf{false} & print eft cross section per channel per order
\\
\tt{with\_resummation\_info} : \textbf{false} & info from resummation: true, false
\egfb
\fakecaption\label{tab:inputoutputoptions}
\end{table}
\begin{table}
\bgfb
\multicolumn{2}{|c|}{\textbf{Table~\ref{tab:massesandscalesoptions}: Masses and scales options}}
\\ 
\tt{m\_higgs} : \textbf{125.0} & higgs mass in GeV
\\
\tt{mur} : \textbf{62.5} & mur
\\
\tt{muf} : \textbf{62.5} & muf
\\
\tt{mt\_msbar} : \textbf{162.7} & MSbar top mass
\\
\tt{mt\_msbar\_ref\_scale} : \textbf{162.7} & reference scale for the top mass in MSbar
\\
\tt{mt\_on\_shell} : \textbf{172.5} & On Shell top mass
\\
\tt{mb\_msbar} : \textbf{4.18} & MSbar bottom mass
\\
\tt{mb\_msbar\_ref\_scale} : \textbf{4.18} & reference scale for the bottom mass in MSbar
\\
\tt{mb\_on\_shell} : \textbf{4.92} & On Shell bottom mass
\\
\tt{mc\_msbar} : \textbf{0.986} & MSbar charm mass
\\
\tt{mc\_msbar\_ref\_scale} : \textbf{3.0} & reference scale for the charm mass in MSbar
\\
\tt{mc\_on\_shell} : \textbf{1.67} & On Shell charm mass
\\
\tt{top\_scheme} : \textbf{msbar} & msbar or on-shell
\\
\tt{bottom\_scheme} : \textbf{msbar} & msbar or on-shell
\\
\tt{charm\_scheme} : \textbf{msbar} & msbar or on-shell
\\
\tt{y\_top} : \textbf{1.0} & factor multiplying the Yt. Set to zero to remove the top quark
\\
\tt{y\_bot} : \textbf{1.0} & factor multiplying the Yb. Set to zero to remove the bottom quark
\\
\tt{y\_charm} : \textbf{1.0} & factor multiplying the Yc. Set to zero to remove the charm quark
\\
\tt{gamma\_top} : \textbf{0.0} & width of top quark
\\
\tt{gamma\_bot} : \textbf{0.0} & width of bottom quark
\\
\tt{gamma\_charm} : \textbf{0.0} & width of charm quark
\egfb
\fakecaption\label{tab:massesandscalesoptions}
\end{table}
\begin{table}
\bgfb
\multicolumn{2}{|c|}{\textbf{Table~\ref{tab:numericalprecisionoptions}: Numerical precision options}}
\\ 
\tt{epsrel} : \textbf{0.0001} & cuba argument: target relative error
\\
\tt{epsabs} : \textbf{0.0} & cuba argument: target absolute error
\\
\tt{mineval} : \textbf{50000} & cuba argument: minimum points to be evaluated
\\
\tt{maxeval} : \textbf{50000000} & cuba argument: maximum points to be evaluated
\\
\tt{nstart} : \textbf{10000} & cuba argument: number of points for first iteration
\\
\tt{nincrease} : \textbf{1000} & cuba argument: number of points for step increase
\\
\tt{cuba\_verbose} : \textbf{0} & cuba argument: verbosity level: 0=silent, 2=iterations printed out
\egfb
\fakecaption
\label{tab:numericalprecisionoptions}
\end{table}

\section{Conclusions}
\label{sec:conclusions}
In this article we presented \ihixs, a comprehensive, easy-to-use tool to derive state of the art predictions for the inclusive production probability of a Higgs boson at the LHC.
The source code for our tool can readily be  downloaded from \url{https://github.com/dulatf/ihixs}.

The theoretical basis of our code was published in ref.~\cite{Anastasiou:2016cez} that details the included contributions and sources of uncertainties.
In summary we include perturbative QCD and electro-weak corrections to the gluon fusion production mechanism. 
\ihixs{} is the first public numerical code that includes exact QCD corrections through N$^3$LO in the heavy top quark effective theory.
QCD corrections with exact mass dependence are include through NLO and approximated at NNLO. 
Furthermore, \ihixs{} provides the option to estimate residual uncertainties for example using the prescriptions outlined in ref.~\cite{Anastasiou:2016cez}.
We have reviewed the essential details of the included contributions and the prescriptions to derive uncertainties.

We have demonstrated the information that can be obtained with \ihixs. 
In particular we have derived the currently most precise predictions for the inclusive production probability for Higgs boson in gluon fusion at a hadron collider.
The inclusion of exact N$^3$LO QCD corrections leads to comparatively small modification of the cross section predictions obtained in ref.~\cite{Anastasiou:2016cez} where N$^3$LO cross sections were computed using a threshold expansion. As a phenomenogical result, we have presented state-of-the art predictions for the Higgs production cross section at the LHC at different collider energies.

Furthermore, we have provide a detailed manual for the usage of \ihixs{}, explaining how the output of our code has to be interpreted and which information can be extracted.
We have discussed how the many features of our tool can be exploited by using a simple runcard system.

The many features of~\ihixs{} provide the user with the ability to perform exhaustive studies of the inclusive Higgs boson production cross section at the LHC and to compare state-of-the-art theoretical predictions to measurements by ATLAS and CMS.
\section*{Acknowledgements}
We would like to thank Babis Anastasiou, Claude Duhr, Elisabetta Furlan, Franz Herzog and Thomas Gehrmann for many useful discussions.
B.M. is supported by the European Comission through the ERC Consolidator Grant HICCUP (No. 614577).
FD is supported by the U.S.~Department of Energy (DOE) under contract DE-AC02-76SF00515.

\appendix
\section{How to add a parameter/option to \ihixs{} }
Adding an input option to {\tt ihixs} requires editing a source file and, hence, re-compilation, but it is fairly straightforward. The file to be edited is 

{\tt src/tools/user\_interface.cpp}. 

The user has to edit the function {\tt UserInterface::UserInterface()} around line 30 of the file. At any empty line within this function, one can add 
\begin{lstlisting}[language=C++]
options.push_back(UIOption(
    "Etot", //: parameter name
    "COM energy of the collider in GeV", //: explanation of what the parameter is. This appears in the --help output
     "Required", //: whether a value is required  for this parameter or not
     "13000.0", //: the default value
     "operational options" //: a classification code
));
\end{lstlisting}
The semantics of the different entries are explained by the inline comments above. The last entry, in particular, determines how the option is displayed in the listing of options produced by {\tt --help}. If the
classification code is one of \texttt{operational options}, \texttt{input-output options}, \texttt{masses and scales options} or \texttt{numerical precision options} then the new option is 
displayed together with the other options in the same group, otherwise it is not displayed at all. Other than that the last entry in the argument list has no effect. 

The third entry in the argument list above is a string that can be either {\tt "Required"} or {\tt "Optional"}. If the option is a parameter, i.e. it holds a value, then the {\tt Required} string should 
be used. If the parameter is a flag, like {\tt --help}, then {\tt Optional} can be used. In such a case a value is not required when defining the option. For example, {\tt Etot} is a parameter with a {\tt "Required"} value, and as a result 
\begin{lstlisting}
    ihixs --Etot
\end{lstlisting}
leads to a runtime error (the program exits indicating that you have invoked {\tt Etot} without specifying a value. On the other hand 
\begin{lstlisting}
    ihixs --help
\end{lstlisting}
works as expected, i.e. one does not need to type {\tt ihixs --help true}.

After re-compiling, the option can be seen with {\tt --help} and can be set either in command line or at the runcard. It can then be used within the program, wherever the {\tt UserInterface} object called {\tt UI} is accessible. If the parameter {\tt foo} is defined and the user wants to access it as a double, he can use 
\begin{lstlisting}[language=C++]
UI.giveDouble("foo")
\end{lstlisting}
wherever in the code {\tt UI} is accessible. There are also functions that case the value to bool ({\tt giveBool("foo")}), integers ({\tt giveInt("foo")}) and strings ({\tt giveString("foo")}) available.

\subsection{How to modify the Wilson Coefficients used in the effective field theory cross section}

The Wilson Coefficients for the gluon-gluon-Higgs effective vertex used in {\tt ihixs} are the ones corresponding to the Standard Model. If the user wants to modify them, to compute the gluon fusion cross section in another model, in which there are heavy particles that modify the gluon-gluon-Higgs effective vertex, the user has to edit three files. The first modification takes place at {\tt src/core/input\_parameters.cpp}
within the function 
\begin{lstlisting}[language=C++]
    void InputParameters::SetUpWilsonCoefficient(const UserInterface &UI)\end{lstlisting}
which is around line 65 of the file. This function has access to the {\tt UI} object and therefore to all user-defined options. Instead of calling 
\begin{lstlisting}[language=C++]
_wc.Configure(_log_muf_over_mt_sq_for_WC,_model.top.scheme());
\end{lstlisting}
the user should call a newly defined function with whatever arguments are necessary for the computation of the modified Wilson Coefficients, e.g.
\begin{lstlisting}[language=C++]
_wc.ModifiedConfigure(UI.giveDouble("myparameter1"), UI.muf, UI.giveBool("myflag"));
\end{lstlisting}

Next, this function has to be declared at\\
{\tt src/higgs/effective\_theory/wilson\_coefficients.h}
and defined at \\
{\tt src/higgs/effective\_theory/wilson\_coefficients.cpp}
similarly to the Standard Model {\tt WilsonCoefficient::Configure} function in that file. Note that the Wilson Coefficients are normalized such that the leading order coefficient $c_0 = 1$. Also note that one should not forget to define the {\tt AsSeries} object {\tt \_c} by using 
\begin{lstlisting}[language=C++]
_c = AsSeries(1,_c0,_c1,_c2,_c3);
\end{lstlisting}
before the end of the {\tt ModifiedConfigure} function.



\clearpage
\bibliographystyle{elsarticle-num}
\bibliography{biblio}
%
\end{document}